\documentclass{aa}  

\usepackage{amsmath}
\usepackage{graphicx}
\usepackage{url}
\usepackage{txfonts}
\def\msun{{\rm ~M}_{\odot}}
\def\rsun{{\rm ~R}_{\odot}}

\def\gpy{{\rm ~Gpc}^{-3} {\rm ~yr}^{-1}}
\def\kms{{\rm ~km} {\rm ~s}^{-1}}
\def\mpy{{\rm ~M}_{\odot} {\rm ~yr}^{-1}}
\def\Mpy{{\rm ~Myr}^{-1}}

\usepackage{color}

\begin{document}

   \title{Binary neutron star formation and the origin of GW170817}

   \author{K. Belczynski\inst{1}\thanks{chrisbelczynski@gmail.com}
          \and
          T. Bulik\inst{2}
          \and
          A.Olejak\inst{2}
          \and
          M. Chruslinska\inst{3}
          \and
          N. Singh\inst{2}
          \and
          N. Pol\inst{4}
          \and
          L. Zdunik\inst{1}
          \and 
          R. O'Shaughnessy\inst{5}
          \and
          M. McLaughlin\inst{4}
          \and
          D. Lorimer\inst{4} 
          \and
          O. Korobkin\inst{6}
          \and
          E.P.J. van den Heuvel\inst{7}
          \and 
          M.B. Davies\inst{8}
          \and
          D. Holz\inst{9}
   }

   \institute{Nicolaus Copernicus Astronomical Center, Polish Academy of Sciences,
           ul. Bartycka 18, 00-716 Warsaw, Poland
         \and
           Astronomical Observatory, Warsaw University, Ujazdowskie 4, 
           00-478 Warsaw, Poland
         \and   
           Department of Astrophysics/IMAPP, Radboud University, P.O. Box 9010, 6500 GL
           Nijmegen, The Netherlands
         \and
           Department of Physics and Astronomy, West Virginia University, Morgantown, 
           West Virginia 26506, USA;\\
           Center for Gravitational Waves and Cosmology, West Virginia University, 
           Chestnut Ridge Research Building, Morgantown, West Virginia 26505, USA
         \and
           Center for Computational Relativity and Gravitation, Rochester
           Institute of Technology, Rochester, New York 14623, USA
         \and
           Center for Theoretical Astrophysics, Los Alamos National
           Laboratory, Los Alamos, NM 87545, USA
         \and
           Astronomical Institute Anton Pannekoek, University of Amsterdam, 
           PO Box 94249, 1090 GE Amsterdam, the Netherlands     
         \and
           Lund Observatory, Box 43, SE-221 00 Lund, Sweden
         \and
            Enrico Fermi Institute, Department of Physics, Department of
            Astronomy and Astrophysics, and Kavli Institute for Cosmological Physics,
            University of Chicago, Chicago, IL 60637, USA 
    }

   \date{Received Dec 25, 2018; accepted ???}

 
  \abstract
{The first neutron star-neutron star merger (NS-NS: GW170817) was detected in
gravitational waves by LIGO/Virgo in a galaxy in which the majority of star 
formation was taking place a long time ago ($\sim 11$ Gyr). Based on this
single event, LIGO/Virgo estimated that local cosmic NS-NS merger rate is 
$110$--$3840\gpy$ ($90\%$ confidence range). Only some extreme evolutionary 
models (with very small NS natal kicks and very high common envelope 
efficiency) can generate NS-NS merger rates in old host galaxies consistent 
with the LIGO/Virgo estimate ($\gtrsim 100 \gpy$). However, we show that 
these models generate rates exceeding empirical Galactic NS-NS merger rates  
based on the large population of Milky Way NS-NS binaries. 

Typically, current evolutionary models produce NS-NS merger rates that are
consistent with the Milky Way empirical rates ($\sim 10$--$200\Mpy$). 
However, these models generate local ($z\approx0$) cosmic NS-NS merger rate 
in old host galaxies ($\sim 1$--$70\gpy$) that are below the LIGO/Virgo 
estimate. The reason behind this tension is the predicted delay time 
distribution between star formation and NS-NS mergers that favors short 
delays. 

Evolutionary models produce a generic steep power-law ($\propto t^{-1}$) 
NS-NS delay time distribution. This limits NS-NS merger rates in old host
galaxies. However, we show that such distribution is consistent 
with observations of Galactic NS-NS binaries; $50\%$ of which show very 
long merger times (much longer than Hubble time). Once model distributions 
are convolved with continuous prolonged ($10$ Gyr) star formation in the 
Galactic disk, then $\sim 20$--$70\%$ (depending on a model) of the 
predicted NS-NS population has very long current Galactic merger times 
($>30$ Gyr). Although NS-NS binaries are formed predominantly with short 
delay times, many of short delay time systems merge and do not make it to 
the present, while long delay time systems survive and contribute to the 
current Galactic NS-NS population. 

This study highlights the tension between the current evolutionary predictions 
and the observation of the first NS-NS merger in an old host galaxy. It is 
crucial to understand that models need to explain not only the LIGO/Virgo rate 
estimate, but also the merger site. 
}


   \maketitle

\section{Introduction}
\label{sec.1}

LIGO/Virgo have discovered the first NS-NS merger through gravitational waves and 
estimated the local cosmic NS-NS merger rate: $110$--$3840\gpy$ ($90\%$ credible 
range with peak probability of $\sim 1000\gpy$; \cite{Abbott2017b,LIGO2018a}). 
The merger was quickly localized in a nearby host galaxy $40$ Mpc away
~\citep{Abbott2017c}. In this galaxy, NGC 4993, star formation peaked $\sim 11$ 
Gyr ago (at level $\sim 10\mpy$) and then was exponentially declining, leading to 
a very low current star formation rate ($\sim 0.01\mpy$). The total mass formed 
in stars was estimated at $7.9 \times 10^{10}\msun$. In particular, the $50\%$ of 
stars formed by $11.2$ Gyr ago, while $90\%$ of stars formed by $6.8$ Gyr ago
~\citep{Blanchard2017}. Stars in NGC 4993 appear to have near solar chemical 
composition~\citep{Blanchard2017,Troja2017,Palmese2017}. 

It was reported that the central parts of NGC 4993 appear to have shell and dust 
structures that may be indicative of a recent minor galaxy merger or mergers. 
However, ~\cite{Troja2017}, based on available UV information, concludes that 
there is no ongoing star formation at the NS-NS merger site and argue against young 
($<2$ Gyr) stellar populations in NGC 4993 based on optical spectral analysis. 
~\cite{Palmese2017} use Dark Energy Camera imaging along with detailed 
spectral analysis of available data to estimate the star formation rate specific
to a potential recent minor merger and finds little to no ongoing star formation 
and conclude that GW170817 is not likely associated with recent star formation. 
Finally, \cite{Blanchard2017} present detailed spectral, photometric and 
image analysis to calculate the star formation history in NGC 4993. In fact,
this analysis reveals some extra component in the recent star formation history
indicative of a minor merger (see the flattening at $t=0.1$--$1$ Gyr ago in their 
exponentially declining SFR; their Fig.3 left-bottom panel). However, this
recent episode provides only a small fraction of the overall stellar mass in NGC
4993. For example, the entire episode ($t=0.1$--$1$ Gyr ago) provides only 
$\sim 0.3 \times 10^9 \msun$ ($0.4\%$ of total mass formed in stars), while 
the most recent part of extra star formation ($t=0.1$--$0.3$ Gyr ago) provides 
only $\sim 0.3 \times 10^8 \msun$ ($0.04\%$ of total mass formed in stars; 
these values can be read of Fig.3 of \cite{Blanchard2017}). 

The discovery of GW170817, accompanied by an unusual weak/offaxis short gamma-ray 
burst (GRB), and by a strong kilonova IR/optical counterpart, and long-lived
X-ray afterglow triggered a search for similar events in the existing data. 
\cite{Troja2018} have identified one such event: GRB 150101B at redshift 
$z=0.1341$. This event is a low-luminosity short GRB, with strong optical 
emission and long-lived X-ray emission estimated to be viewed at an angle of
$13$ degrees. At this distance ($\sim 600$ Mpc) a NS-NS merger is undetectable 
by LIGO/Virgo (even at their design sensitivity). However, the physical
properties seem to by symptomatic of a NS-NS merger, or at least similar to 
the one observed in NGC 4993. GRB 150101B is located within its host galaxy
$7.3$ kpc from its center. The most interesting fact is that this is an
early type galaxy with mean stellar age of $2^{+6}_{-1}$ Gyr typical of 
elliptical galaxies (see Fig.5 of \cite{Troja2018}). 

Isolated binary evolution in galactic fields and dynamical evolution in 
globular or nuclear clusters are the main formation channels for NS-NS binaries. 
For example, in the Milky Way there are $18$ known NS-NS systems; $16$ of which 
are found in the Galactic disk/field, and $2$ in Galactic globular clusters 
(see Tab.~\ref{tab.obs}). The GW170817 projected distance from the center of 
NGC 4993 is $2.1$ kpc: within the galaxy half-light radius. Also there is no 
visible globular cluster in the vicinity of the merger~\citep{Blanchard2017,
Troja2017,Palmese2017}. Evolutionary predictions indicate that binary 
evolution dominates the formation rate of NS-NS mergers at late times after 
star formation in galaxies like NGC 4993, over globular and nuclear cluster 
rates by $2$--$3$ orders of magnitude~\citep{Belczynski2018}. 

In this study, we will focus on the isolated binary evolution channel in the context 
of the formation of GW170817 in NGC 4993. Current evolutionary predictions
based on population synthesis calculations typically generate local cosmic
($z\approx0$) NS-NS merger rates at the level $\sim 100 \gpy$. However, if 
several not fully constrained evolutionary parameters (e.g., NS natal kicks, 
common envelope efficiency, Roche lobe overflow treatment) are pushed in 
favor of NS-NS formation, the merger rates can reach $\sim 500\gpy$ 
~\citep{Chruslinska2018,Kruckow2018,Vigna2018,Mapelli2018}\footnote{There is 
one exception to this general consensus. Apparently, simulations based on BPASS 
populations synthesis code~\citep{Eldridge2017} generate very large NS-NS merger 
rates: $1000$--$5000\gpy$ if BPASS newly proposed NS natal kick prescription is
applied~\citep{Bray2018,Eldridge2018}. However, note that such high rates are 
not consistent with empirical Galactic NS-NS merger rate estimates (see Sec.
~\ref{sec.5} for discussion).}. This is consistent  with the 
LIGO/Virgo low-end of $90\%$ confidence range of the NS-NS merger rate
($110$--$3840\gpy$). It may seem like there is apparently no tension between
the LIGO/Virgo observation and these theoretical predictions. However, the merger 
rate of NS-NS systems is directly proportional to star forming mass for 
isolated binary evolution and evolutionary predictions show a generic delay
time distribution ($\propto t^{-1}$) that favors short delay times.
This indicates that NS-NS mergers are more likely in host galaxies with
ongoing or recent star formation. If we take into account the fact that locally 
only $1/3$ of galaxies are ellipticals~\citep{Conselice2016} then predicted 
isolated binary evolution rates drop from $\sim 500\gpy$ to $\sim 170\gpy$, 
which is then only marginally consistent with the LIGO/Virgo empirical estimate. 
Note that such estimate completely ignores the fact that NS-NS merger rate depends 
strongly on delay time distribution and thus star formation history in a given 
type of host galaxy. This factor can be easily assessed for elliptical hosts. 
Elliptical galaxies typically formed the majority of stars $1$--$10$ Gyr ago
~\citep{Gallazzi2006}. There is an expectation (confirmed by evolutionary 
calculations; e.g., see Sec.~\ref{sec.41}) that NS-NS binaries typically 
begin merging $\sim 100$ Myr after star formation (stellar evolution takes 
several tens of Myr to form NSs out of massive stars). If the above is taken
into account, then the current NS-NS merger rate in ellipticals decreases 
by $1$--$2$ orders of magnitude with respect to the above optimistic estimate. 
This means that the current local cosmic merger rate in ellipticals predicted 
by evolutionary calculations is at most $\sim 1.7$--$17 \gpy$. If the LIGO/Virgo 
{\em single} observation is to be trusted (note that small number statistics/Poisson 
errors were taken into account in the LIGO/Virgo rate estimate) then there is 
a tension between evolutionary predictions and the LIGO/Virgo observation of 
GW170817. In other words, current evolutionary predictions cannot explain formation 
of NS-NS mergers in old host galaxies at such rates that it would warrant 
detection at the current LIGO/Virgo sensitivity ($\sim 70$ Mpc; 
\cite{Abbott2017b}), as one expects a merger in elliptical galaxy to be 
detected every 50-500 years. 

We can use the same line of reasoning to show that a potential minor galaxy 
merger in the recent past of NGC 4993 was not likely to produce a NS-NS merger 
through isolated binary evolution. Let's start with all elliptical galaxies 
in the local Universe; their entire star forming mass is predicted to produce  
a NS-NS merger rate of $\sim 170\gpy$ in optimistic evolutionary models that 
assume constant star formation. Let's put $1\%$ of this star forming mass at 
such preferable times that all NS-NS mergers could be detected by the LIGO/Virgo. 
This way we circumnavigate the issue of long delays expected for elliptical host 
galaxies. Then the expected current merger rate is only $\sim 1.7\gpy$, and in 
clear tension with LIGO/Virgo empirical estimate. Note that for this exercise we 
assumed a rather large stellar mass of a minor merger (for NGC 4993 we estimated 
this mass to be only $0.04-0.4\%$ of entire galaxy star forming mass) and we 
assumed that every elliptical in the LIGO/Virgo range went through such minor
merger at the best preferable time so all NS-NS mergers could have been detected. 

It seems like the steep delay time distribution is the major factor 
limiting NS-NS merger rates in old host galaxies in evolutionary models. 
In the following sections we will re-examine the issue of delay time
distributions obtained in evolutionary models and compare them with 
available observations of Galactic NS-NS binaries. If models with steep 
delay time distributions do not reproduce the properties of Galactic 
NS-NS systems (their observed merger times and merger rates) then this 
will indicate a need for a major revision of models. However, if the models 
do reproduce the observed population of Galactic NS-NS binaries, then  
other options may need to be invoked to explain GW170817. For example, 
GW170817 was a significant statistical fluctuation, or this NS-NS merger 
was produced by other process than isolated binary evolution.

\begin{table*}
\caption{Galactic NS-NS binaries\tablefootmark{a}} 
\centering          
\begin{tabular}{l| c c c c c c c c}     
\hline\hline       
Name                     &     type & $M_{\rm psr}$\tablefootmark{b} & $M_{\rm com}$ & $P_{\rm orb}$ &       $a$ & $e$    & $t_{\rm mer}$\tablefootmark{c} & reference\tablefootmark{e} \\
                         &          &     [$\msun$] &     [$\msun$] &         [day] & [$\rsun$] &        &         [Gyr] & \\ 
\hline
\hline
{\bf field:}             &          &               &               &               &           &        &               &     \\
1)  J1946+2052           & recycled &          1.25 &          1.25 &         0.076 &     1.028 &  0.06  &         0.042 & [1] \\
2)  J1757-1854           & recycled &          1.34 &          1.39 &         0.183 &     1.897 &    0.6 &         0.079 & [2] \\ 
3)  J0737-3039           &    young &         1.338 &         1.249 &         0.102 &     1.261 &  0.088 &         0.085 & [3,4,5] \\
4)  B1913+16             & recycled &         1.440 &         1.389 &         0.323 &     2.801 &  0.617 &         0.301 & [6,7] \\
5)  J1906+0746           &    young &         1.291 &         1.322 &         0.166 &     1.750 &  0.085 &         0.308 & [8,9] \\
6)  J1913+1102           & recycled &          1.64 &          1.25 &         0.206 &     2.090 &   0.08 &         0.473 & [10,11] \\
7)  J1756-2251           & recycled &         1.341 &         1.230 &         0.320 &     2.696 &  0.181 &         1.660 & [12,13] \\ 
8)  B1534+12             & recycled &         1.333 &         1.346 &         0.421 &     3.282 &  0.274 &         2.736 & [14] \\
                         &          &               &               &               &           &        &               & \\ 
9)  J1829+2456           & recycled &         1.295 &         1.295 &         1.176 &     6.436 &  0.139 &         55.36 & [15] \\
10) J1411+2551           & recycled &         1.265 &         1.265 &          2.61 &      10.9 &   0.16 &         471.3 & [16] \\
11) J0453+1559           & recycled &         1.559 &         1.174 &         4.072 &      15.0 &  0.113 &         1,452 & [17] \\
12) J1811-1736           & recycled &         1.285 &         1.285 &        18.779 &      40.7 &  0.828 &         1,794 & [18] \\   
13) J1518+4904           & recycled &         1.359 &         1.359 &         8.634 &      24.7 &  0.249 &         8,853 & [19] \\
14) J1755-2550           &    young &           1.3 &           1.3 &         9.696 &      26.3 &  0.089 &        15,917 & [20,21] \\
15) J1753-2240           & recycled &           1.3 &           1.3 &        13.638 &      33.0 &  0.304 &        28,646 & [22] \\
16) J1930-1852           & recycled &         1.295 &         1.295 &        45.060 &      73.1 &  0.399 &       531,294 & [23] \\ 
\hline
{\bf globular clusters:} &          &               &               &               &                    &               &         \\ 
17) B2127+11C            & recycled &         1.358 &         1.354 &         0.335 &     2.830 &  0.681 &         0.217 & [24,25] \\ 
                         &          &               &               &               &                    &               &         \\ 
18) J1807-2500B\tablefootmark{d} & recycled & 1.366 &         1.206 &         9.957 &      26.7 &  0.747 &         1,044 & [26]    \\         
\hline      
\hline
\end{tabular}
\label{tab.obs}
\tablefoot{ \\
\tablefootmark{a}{All known Galactic systems}\\
\tablefootmark{b}{If only total mass is reported in literature, we use 
$M_{\rm psr}$=$M_{\rm com}$=$0.5 M_{\rm tot}$.}\\
\tablefootmark{c}{Current time to merger estimated from NS masses and
currently measured orbital parameters: $a$ and $e$.}\\
\tablefootmark{d}{This may be  potentially a NS-WD system:
\url{https://www3.mpifr-bonn.mpg.de/staff/pfreire/NS_masses.html}.}\\
\tablefootmark{e}{[1]: \citet{Stovall18}; [2]: \citet{Cameron18}; [3]: \citet{Kramer06}; 
[4]: \citet{Breton08}; [5]: \citet{Ferdman13}; [6]: \citet{HulseTaylor75}; [7]: \citet{Weisberg10};
[8]: \citet{Lorimer06}; [9]: \citet{vanLeeuwen15}; [10]: \citet{Lazarus16}; [11]: \citet{Ferdman18IAU}; 
[12]: \citet{Faulkner05}; [13]: \citet{Ferdman14}; [14]: \citet{Fonseca14}; [15]: \citet{Champion04}; 
[16]: \citet{Martinez17}; [17]: \citet{Martinez15}; [18]: \citet{Corongiu07}; [19]: \citet{Janssen08};
[20]: \citet{Ng15}; [21]: \citet{Ng18}; [22]: \citet{Keith09}; [23]: \citet{Swiggum15}; 
[24]: \citet{Anderson90}; [25]: \citet{Jacoby06}; [26]: \citet{Lynch12}.}
}
\end{table*}

\section{Binary evolutionary models}

Binary evolution calculations are performed with the upgraded population
synthesis code {\tt StarTrack} \citep{Belczynski2002,Belczynski2008a}.
The existing improvements relevant for massive star evolution include
updates to the treatment of common envelope (CE) evolution \citep{Xu2010,Dominik2012}, 
the compact object masses produced by core collapse/supernovae \citep{Fryer2012,Belczynski2012}
with the effect of pair-instability pulsation supernovae and pair-instability
supernovae \citep{Woosley2007,Belczynski2016c}, stellar binary initial conditions set by
observations \citep{Sana2012,deMink2015}, and observationally constrained star formation
and metallicity evolution over cosmic time \citep{Madau2014,Belczynski2016b}.
The code adopts by default the fallback-decreased natal kick prescription 
\citep{Belczynski2017b}.

For our study we select fifteen evolutionary models (M10 and NN1--NN14) which 
differ significantly by some evolutionary assumptions important for massive star 
evolution. The model M10 is fully described by \cite{Belczynski2017b}.
For model M10 standard evolutionary assumptions are adopted: standard NS/BH
masses~\citet{Fryer2012} with pair-instability pulsations and SNe, 
low-to-no BH natal kicks (set by fallback), high kicks for core-collapse (CC) NSs 
drawn from Maxwellian with with 1-dimensional $\sigma=265\kms$ and modified by 
fallback, no natal kicks for electron-capture supernova (ECS) NS formation, $50\%$ 
non-conservative RLOF, $10\%$ Bondi-Hoyle rate accretion onto NS/BH in CE, no 
effects of rotation on stellar evolution\footnote{Binary component spins are 
followed (tides, magnetic braking and change of inertia). However, stellar 
rotation does not alter internal star properties (He/CO core mass).}, initial 
binary parameters from \citet{Sana2012}, and massive star winds 
\citet{Vink2001,Belczynski2010b} with LBV winds calibrated to produce BHs with 
maximum mass of $15\msun$ at current Galactic disk metallicity 
($Z=0.02$: $(dM/dt)_{\rm LBV} = 1.5 \times 10^{-4} \mpy$).

In model NN2 we assume high CC NS natal kicks with $\sigma=265\kms$ without 
any fallback effect. ECS NSs receive exactly the same natal kicks as CC NSs.
During CE we assume that $100\%$ of orbital energy is used to eject the envelope,
and this corresponds to CE efficiency of $\alpha_{\rm CE}=1.0$. During 
non-conservative RLOF, $20\%$ of donor mass transfer is accreted onto
non-degenerate companion, while $80\%$ of donor mass transfer is ejected from 
binary.   

In model NN14, all the assumptions are the same as in model NN2, with one
exception. We assume here that CC NS natal kicks are moderated by small amount 
of fallback expected even in NS formation~\citep{Fryer2012}. All NS natal 
kicks are thus somewhat smaller than in model NN2. 

In model NN3, all the assumptions are the same as in model NN14, with one
exception. We assume here that ECS NSs receive different natal kicks than CC
NSs. Natal kicks for ECS NSs in this model are assumed to be zero. 

In models NN7 and NN8 we assume similar physics as in model NN2, but we
introduce two modifications. CC NS natal kicks are drawn from the
distribution with smaller kicks: $\sigma=133\kms$. ECS NSs are assumed to
receive different natal kicks than CC NSs: drawn from Maxwellian with
1-dimensional $\sigma=66\kms$ for model NN7, and with zero natal kicks for
model NN8.

In model NN11 we use the same assumptions as in model NN2, however we lower
significantly CC NS natal kicks to $\sigma=66\kms$. Note that this applies
to all NSs, as ECS natal kicks are treated the same way as the CC NS kicks. 

In models NN9 and NN10 we assume similar physics as in model NN11, but we
allow for different ECS NS natal kicks. ECS natal kicks are drawn from 
Maxwellian with 1-dimensional $\sigma=33\kms$ for model NN9, and with zero 
natal kicks for model NN10.

In models NN12 and NN4 we use the same assumptions as in models NN2 and NN11. 
However, we further lower CC NS natal kicks to $\sigma=33\kms$ for model
NN12 and to $\sigma=0\kms$ (no natal kicks) for model NN4. 

In model N13, we adopt same assumptions as in model NN4, with one modification 
of CE treatment: very high common envelope efficiency $\alpha_{\rm CE}=10$. 
Note that this model has not only the lowest possible NS natal kicks (like in 
NN4), but also most likely unrealistically high CE efficiency. 

In models NN1, NN5 and NN6 we adopt ~\cite{Bray2018} natal kicks for both
ECS and CC NSs (they use the same prescription for CC and ECS NSs).   
The 3-dimensional natal kick magnitude is taken from 
\begin{equation}
v_{\rm kick}=\alpha (M_{\rm ejecta}/M_{\rm remnant}) + \beta, 
\label{eq.be18}
\end{equation}
while the direction of natal kick is random. 
This prescription was adopted with specific natal kick parameters proposed
by these authors:  $\alpha=100\kms$ and $\beta=-170\kms$; note that this
prescription produces almost zero NS natal kicks. Each of these three models 
differs only by one parameter, CE efficiency:  $\alpha_{\rm CE}=0.1,\ 1.0,\ 10$
for NN5, NN1 and NN6, respectively. All the rest of physical assumptions are
the same as in model NN2. 

For main sequence donors CE is assumed always to lead to binary component merger 
aborting binary evolution. In case of evolved donors (beyond Hertzsprung gap) we use 
the energy balance approach of \cite{Webbink1984} with updates on estimates of CE binding 
energy to test whether a given system survives CE or not. It was argued that the 
outcome of CE is not clear in case of Hertzsprung gap donors \citep{Belczynski2007,
Pavlovskii2017}. For all models we either allow for CE survival (based on energy balance) 
through CE with Hertzsprung gap donor (submodels A), or we eliminate such systems from 
our sample (submodels B).

For all models/submodels we have formed NS-NS binaries. We can calculate now
the delay time from star formation to NS-NS merger for each system as:
\begin{equation}
t_{\rm delay} = t_{\rm evol} + t_{\rm mer,i} 
\label{eq.1}
\end{equation}
where $t_{\rm evol}$ is the time from star formation (Zero Age Main Sequence
for both stars in a given binary) to the formation of the NS-NS system, and 
$t_{\rm mer,i}$ is the (intrinsic) merger time: time from NS-NS formation to
final coalescence of two NSs. The evolutionary time is set by the evolution of 
massive stars that form NSs (usually several to several tens of Myr), while
the intrinsic merger time is set by the two NS masses and orbital separation and
eccentricity (\cite{Peters1964}; it can range from 0 to very large values
for wide orbits). In Figures~\ref{fig.tm10}, ~\ref{fig.tnn3}, ~\ref{fig.tnn8}, 
~\ref{fig.tnn13}, ~\ref{fig.tnn1}, and ~\ref{fig.tnn6} we show delay time 
distributions for some of our evolutionary models.

\section{Criteria for a NS-NS binary selection from models}
\label{sec.crit}

\subsection{Milky Way NS-NS merger rates}
\label{sec.crit1}

In each simulation, we evolve $N_{\rm mbin}=2\times 10^6$ massive binaries: 
primary in mass range $5$--$150\msun$ and secondary in mass range $3$--$150\msun$. 
With our adopted binary fraction of $50\%$, and our adopted broken power-law 
initial mass function (IMF) with slope of $-2.3$ for massive stars our total 
simulated mass (extended to hydrogen burning limit: $0.08\msun$) is 
$M_{\rm sim}=2.8\times10^8\msun$. In each simulation we generate some number
of NS-NS binaries with delay time shorter than age of Galactic disk ($10$ Gyr):
$N_{\rm nsns}$. Assuming mass of the Galactic disk to be 
$M_{\rm MW,disk}=5.17\times 10^{10}\msun$, and continuous star formation in
Galactic disk during $10,000$ Myr we can estimate Galactic disk merger rate 
from
\begin{equation}
{\cal R}_{\rm MW}= {M_{\rm MW,disk} \over M_{\rm sim}} {N_{\rm nsns} \over
10,000} Myr^{-1}. 
\label{eq.mwrate}
\end{equation}
Milky Way disk merger rates for various models are reported in 
Tables~\ref{tab.stat10} and ~\ref{tab.rates}.

\subsection{Current Milky Way NS-NS merger times}
\label{sec.crit2}

We assign synthetic binaries that form NS-NS systems Galactic birth time: 
$t_{\rm birth}$ in range $0$--$10$ Gyr; these times are drawn from uniform
distribution that may well represent a continuous star formation rate and age 
of the Galactic disk. Synthetic binaries form NS-NS systems at time: 
$t_{\rm form}=t_{\rm birth}+t_{\rm evol}$. From the NS-NS formation time to 
current Galactic disk age ($10$Gyr) we evolve NS-NS systems according to
loss of angular momentum through emission of gravitational waves. This leads
to a decrease of the orbital parameters $a$ and $e$. We record current Galactic 
time NS-NS orbital properties: $a_{\rm MW}$ and $e_{\rm MW}$ (note that
these values may differ quite significantly from the binary orbital parameters
at the formation of NS-NS system). Note also that some NS-NS may have merged 
before reaching the current Galactic time, and that some binaries may not have 
formed a NS-NS by the current Galactic time (these systems are not included in 
the follow-up analysis):  

\begin{enumerate}

\item the synthetic NS-NS binaries at Galactic current time $t_{\rm galax}=10$ 
Gyr are selected 

\item the NS-NS current merger time is calculated: $t_{\rm mer,MW}$ based on
the current Galactic orbital parameters $a_{\rm MW}$ and $e_{\rm MW}$ 

\item if $t_{\rm mer,MW}>t_{\rm cutoff}=6\times10^5$ Gyr NS-NS is removed: 
      merger time longer than for the longest observed NS-NS system

\item a radio (young) pulsar is defined as a NS with age: 
$t_{\rm age}<\tau_{\rm radio}$ yr, where $\tau_{\rm radio}$ is radio pulsar
lifetime drawn from specific distribution (see Fig.~\ref{fig.radio}) 

\item a recycled pulsar is defined as a NS with age: 
$t_{\rm age}<t_{\rm recyc}$ yr and entire (RLOF+CE+WIND) accreted 
mass: $dM>dM_{\rm rec}=0.1\msun$. $t_{\rm recyc}$ is recycled pulsar   
lifetime drawn from specific distribution (see Fig.~\ref{fig.recyc})

\item NS-NS systems without at least one radio or one recycled pulsar at
current Galactic time are removed (as non-detectable in radio) 

\item the leftover NS-NS binaries are the current Galactic radio-detectable
population. 

\end{enumerate}

The properties of such selected sample of model NS-NS systems are compared 
with the current merger times of the known Galactic field NS-NS binaries: 
see Table~\ref{tab.stat10} and Figures~\ref{fig.m10A}, ~\ref{fig.m10B}, 
~\ref{fig.m11A}, and ~\ref{fig.m11B}.

\subsection{NS-NS merger rate in elliptical galaxies}
\label{sec.crit3}

We use the Illustris cosmological simulation~\citep{Vogelsberger2014,Snyder2015} 
to estimate the mass of all elliptical galaxies within cube with side of $L=100$ 
Mpc. The volume of such cube ($0.001$ Gpc$^3$) approximately corresponds to 
volume in which LIGO/Virgo was able to detect NS-NS mergers in O1/O2 runs. 
Following details discussed in Appendix A of \cite{Belczynski2018} this mass
is $M_{\rm ell,tot}=1.1 \times 10^{14} \msun$. Since our simulation mass is
only $M_{\rm sim}=2.8\times10^8\msun$ we will have to multiply number of
NS-NS binaries that we form in each model by 
$F_{\rm x}=M_{\rm ell,tot}/M_{\rm sim}=3.9\times10^5$.  
This gives us total number of NS-NS binaries formed in all ellipticals
within LIGO/Virgo reach. This number then needs to be modified by adopted
star formation history in elliptical galaxies. We perform two calculations. 
In one we assume that all ellipticals in local Universe are $10$ Gyr old
(e.g., approximately NGC4993), and in the other that ages of ellipticals 
in local Universe are uniformly distributed in range $1$--$11$ Gyr
~\citep{Gallazzi2006}. We assume that star formation in ellipticals was a
burst event: all stars formed at the same time corresponding to the age of 
a given elliptical galaxy. Final current time (present) local rate of NS-NS
mergers (${\cal R}_{\rm ell}$) from elliptical galaxies is estimated with 
the use of NS-NS delay time distribution found in our models (see 
Sec.~\ref{sec.41}). The rates for all the models are listed in Table
~\ref{tab.rates}.

\section{Results}

\subsection{Galactic NS-NS delay time distribution}
\label{sec.41} 

Note that delay times (from ZAMS to merger; eq.~\ref{eq.1}) for NS-NS binaries 
are not known as exact recycled (majority of Galactic population) pulsar ages 
cannot be established. However, the delay time distribution is a primary factor 
(along star forming mass) setting the NS-NS merger rate in any host galaxy. 

\begin{figure}   
\hspace*{-0.3cm}   
\includegraphics[width=9.2cm]{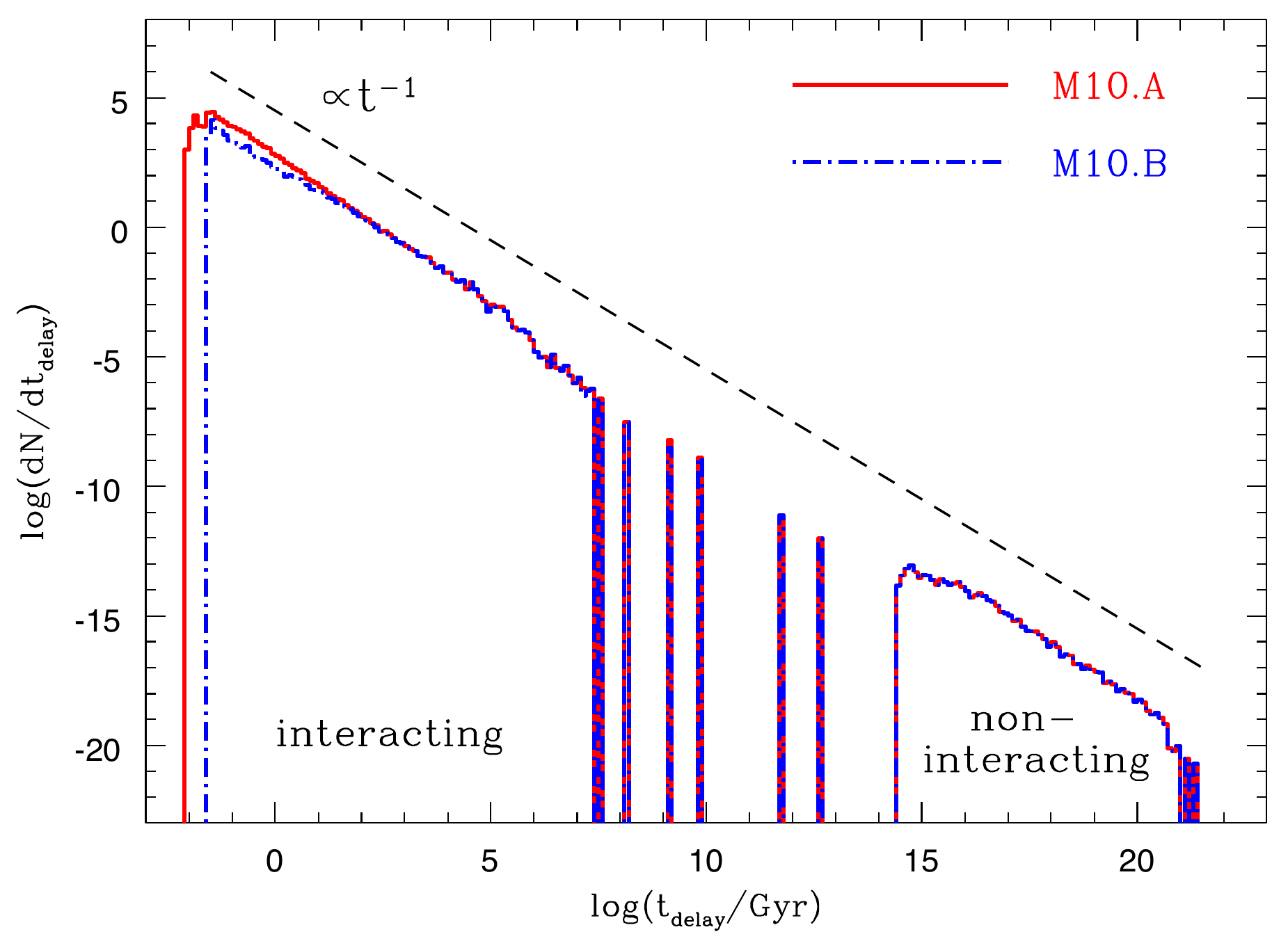}  
\caption{    
Model M10: NS-NS delay time distribution. Note generic power-law
($\propto t^{-1}$) shape of the distribution. We indicate NS-NS 
populations that formed out of interacting (common envelope) and 
non-interacting (no common envelope) binary progenitor systems. 
}
\label{fig.tm10}
\end{figure}

\begin{figure}   
\hspace*{-0.3cm}   
\includegraphics[width=9.2cm]{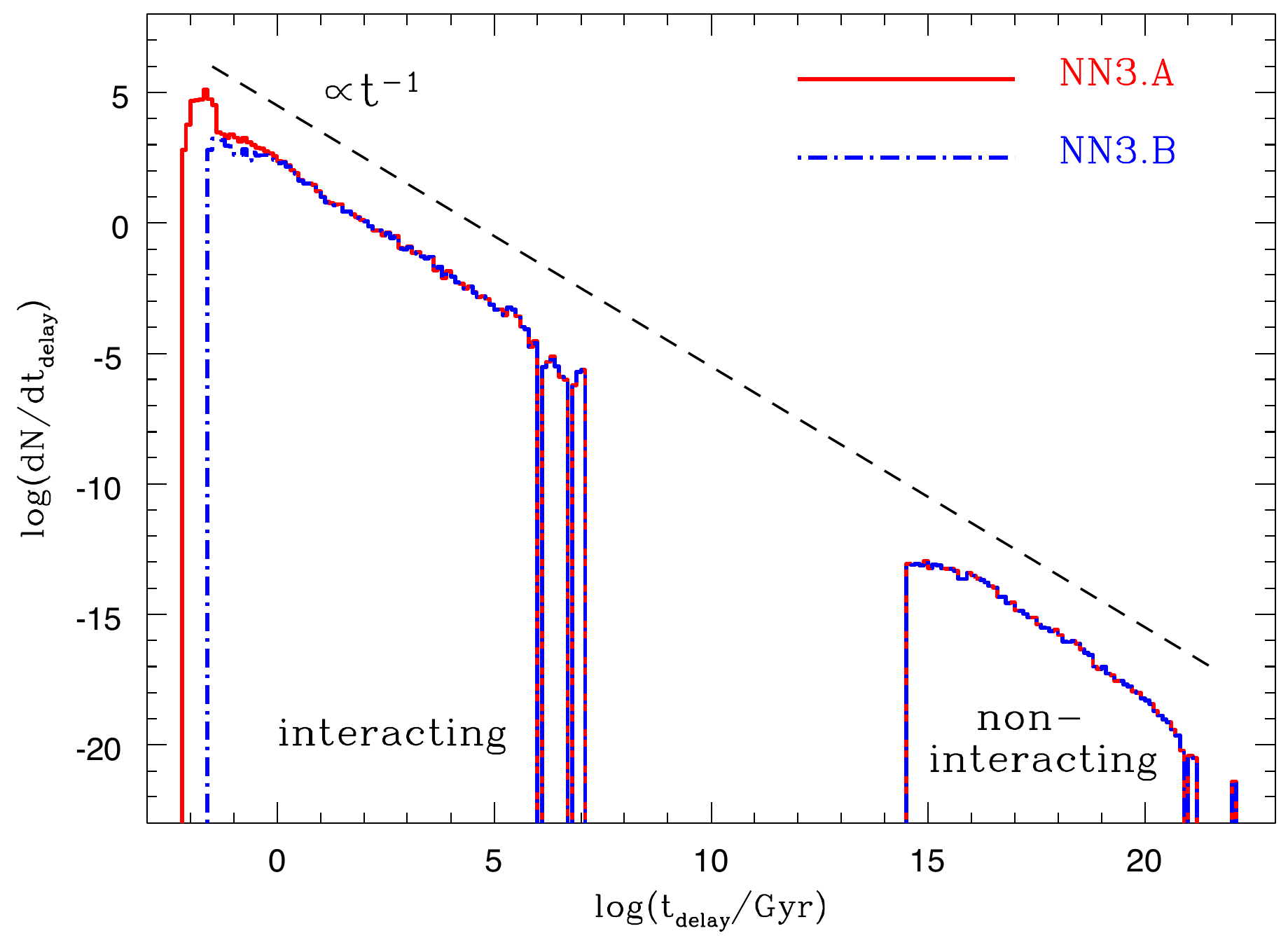}  
\caption{    
Model NN3: NS-NS delay time distribution. Labels same as in 
Figure~\ref{fig.tm10}. 
}
\label{fig.tnn3}
\end{figure}

\begin{figure}   
\hspace*{-0.3cm}   
\includegraphics[width=9.2cm]{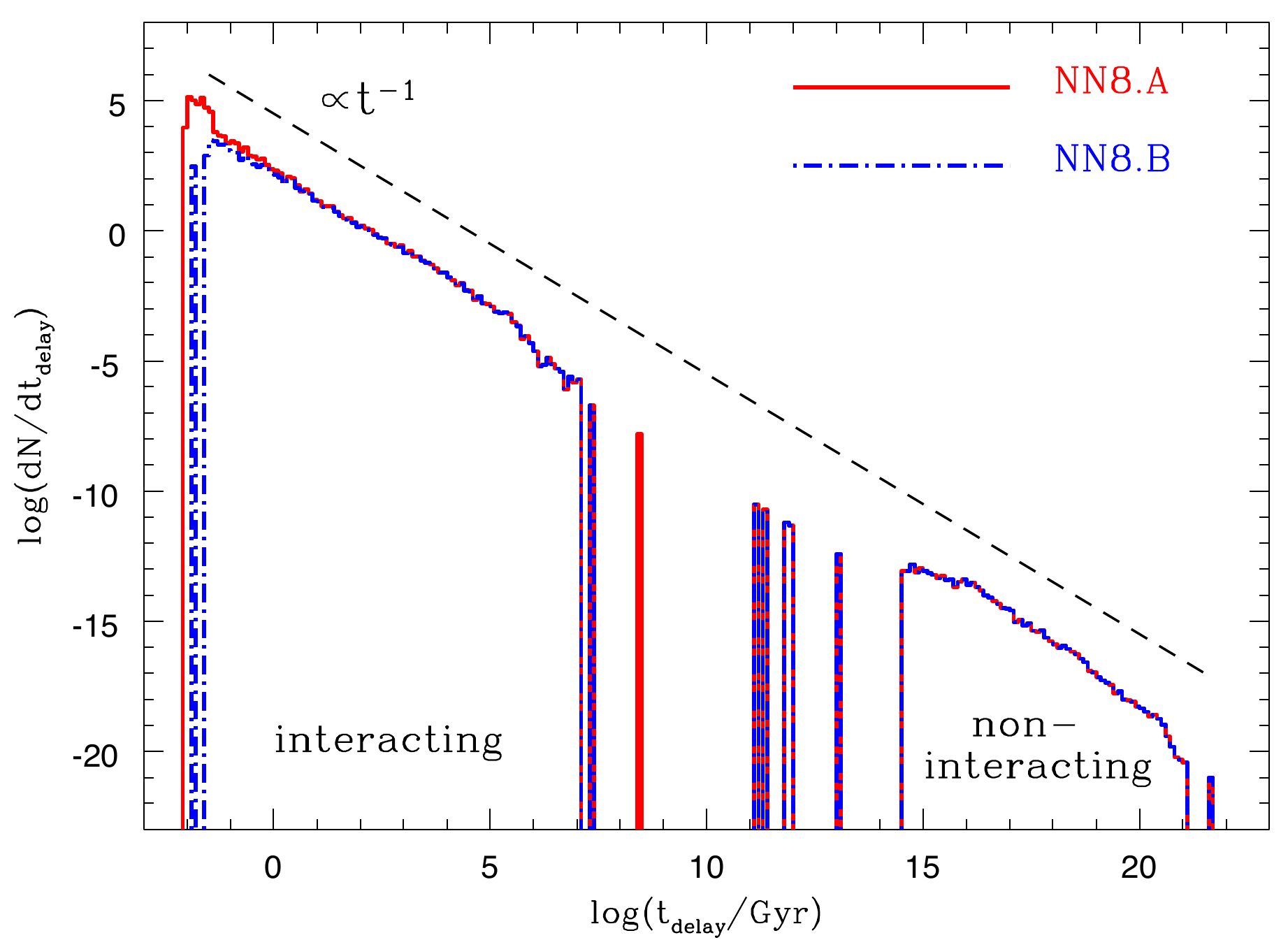}  
\caption{    
Model NN8: NS-NS delay time distribution. Labels same as in 
Figure~\ref{fig.tm10}. 
}
\label{fig.tnn8}
\end{figure}

Figures~\ref{fig.tm10}, ~\ref{fig.tnn3}, and ~\ref{fig.tnn8} show delay time 
distributions for NS-NS binaries in models M10, NN3, NN8. These are typical
models, that adopt high to moderate natal kicks for CC NSs and zero natal
kicks for ECS NSs, non-conservative RLOF with $50$--$80\%$ of mass loss from
binary, and fully efficient CE evolution ($\alpha_{\rm CE}=1$).  
The distributions for all these models (and submodels A and B) are very similar 
and they follow a power-law $t_{\rm delay}^{-1}$ over many orders of magnitude. 
Distributions begin at $\sim 10-100$ Myr and end around $10^{21}$ Gyr. Note the 
gap in the delay time distribution at times $10^{7}$--$10^{15}$ Gyr. This gap 
separates the systems that formed through CE phase (left of the gap; short 
delays) and that did not evolve through CE phase (right of the gap; long delays). 

It is clear that the merger rate of NS-NS systems decreases steeply from the end 
of star formation in a given host galaxy. For example, a galaxy with a burst of 
star formation within last $100$ Myr would have a NS-NS merger rate $100$ times 
larger than an elliptical galaxy of the same mass that had its burst of star 
formation $10$ Gyr ago. A spiral host galaxy that forms stars in continuous fashion 
would generate a merger rate between these two extreme cases. This has two 
consequences.
\begin{enumerate}

\item NGC4993, in which GW170817 was found, had a peak of star formation 
$\sim 11$ Gyr ago~\citep{Blanchard2017} and therefore is a very unlikely 
galaxy to host the first NS-NS merger. The majority of NS-NS mergers are 
predicted to happen quickly after star formation; in starburst or spiral 
galaxies, with only a small fraction occurring in old ellipticals.  
 
\item The evolutionary NS-NS merger rate predictions that are pushed to
reach $300-500\gpy$ \citep{Chruslinska2018,Kruckow2018,Vigna2018,Mapelli2018} 
are indeed consistent with the LIGO/Virgo rate estimate ($110$--$3840\gpy$).
However, these predictions are based on overall star formation in the local 
Universe  in {\em all} sorts of host galaxies. These predictions produce steep 
power-law delay times for NS-NS binaries as in our models described above
(M10, NN3 or NN8). If the rates are recalculated {\em only} for old host 
galaxies (resembling NGC4993), they drop by $\sim 2$ orders of magnitude and 
are thus in tension with the LIGO/Virgo estimate \citep{Belczynski2018}.

\end{enumerate}
 
The power-law shape of the delay time distribution of NS-NS binaries is a generic 
outcome of modern population synthesis predictions and is naturally explained by 
the underlying physics. The initial orbital separation distribution for massive 
O/B stars follows approximately a power-law $\propto a^{-1}$. 
Evolutionary processes in close binary systems (in particular CE phase) reduce 
initial binary separations by $1-2$ orders of magnitude, producing an even steeper 
power-law distribution of separations at NS-NS formation $\propto a^{-3}$.
After NS-NS formation, binary orbit decays due to emission of gravitational 
radiation (GR) at the rate that is firmly established and strongly depends on 
orbital separation of two NSs: $t \propto a^{4}$ \citep{Peters1964}.
Assuming that the distribution of orbital separations at the time of formation 
of the NS-NS binary can be described by a power-law $dN/da \propto a^{-\beta}$, 
we obtain the distribution of the merger times $dN/dt_{merg} \propto t^{-\beta/4-3/4}$.
The exponent only weakly depends on $\beta$ and for $\beta=1$ we obtain
$dN/dt_{merg} \propto t^{-1}$, while for $\beta=3$ we obtain 
$dN/dt_{merg} \propto t^{-1.5}$

\begin{figure}   
\hspace*{-0.3cm}   
\includegraphics[width=9.2cm]{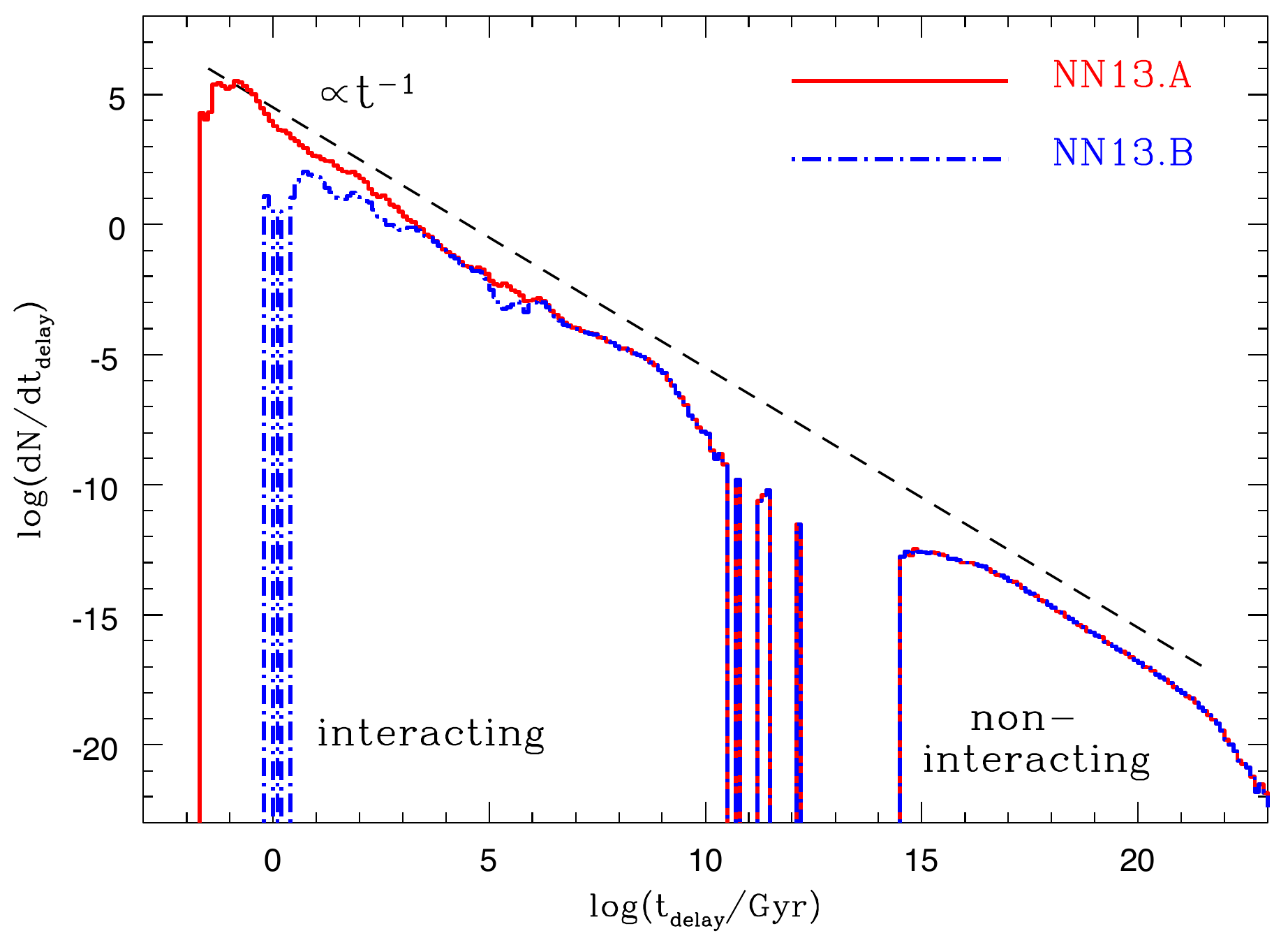}  
\caption{    
Model NN13: NS-NS delay time distribution. Labels same as in 
Figure~\ref{fig.tm10}. 
}
\label{fig.tnn13}
\end{figure}

\begin{figure}   
\hspace*{-0.3cm}   
\includegraphics[width=9.2cm]{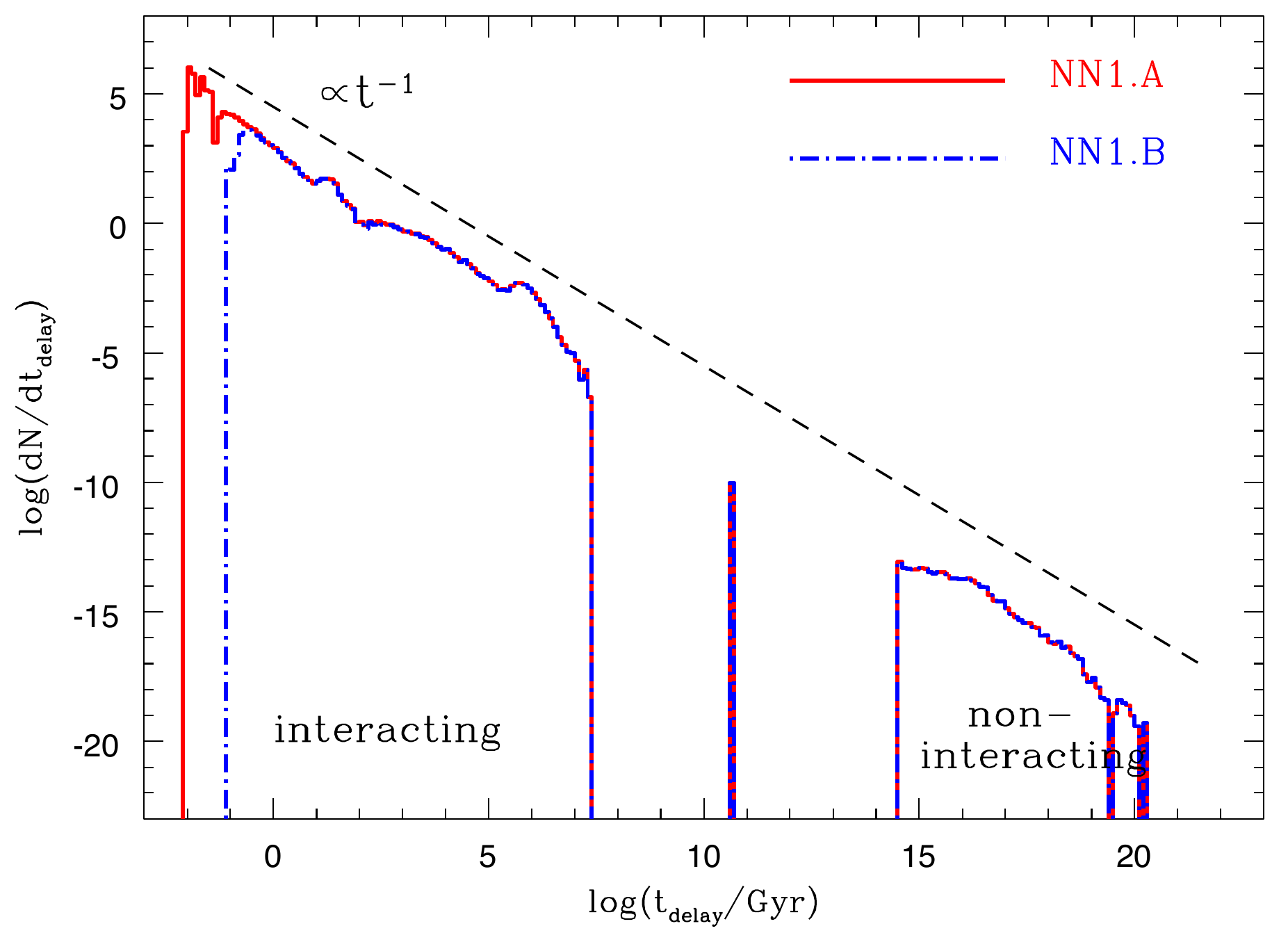}  
\caption{    
Model NN1: NS-NS delay time distribution. Labels same as in 
Figure~\ref{fig.tm10}. 
}
\label{fig.tnn1}
\end{figure}

\begin{figure}   
\hspace*{-0.3cm}   
\includegraphics[width=9.2cm]{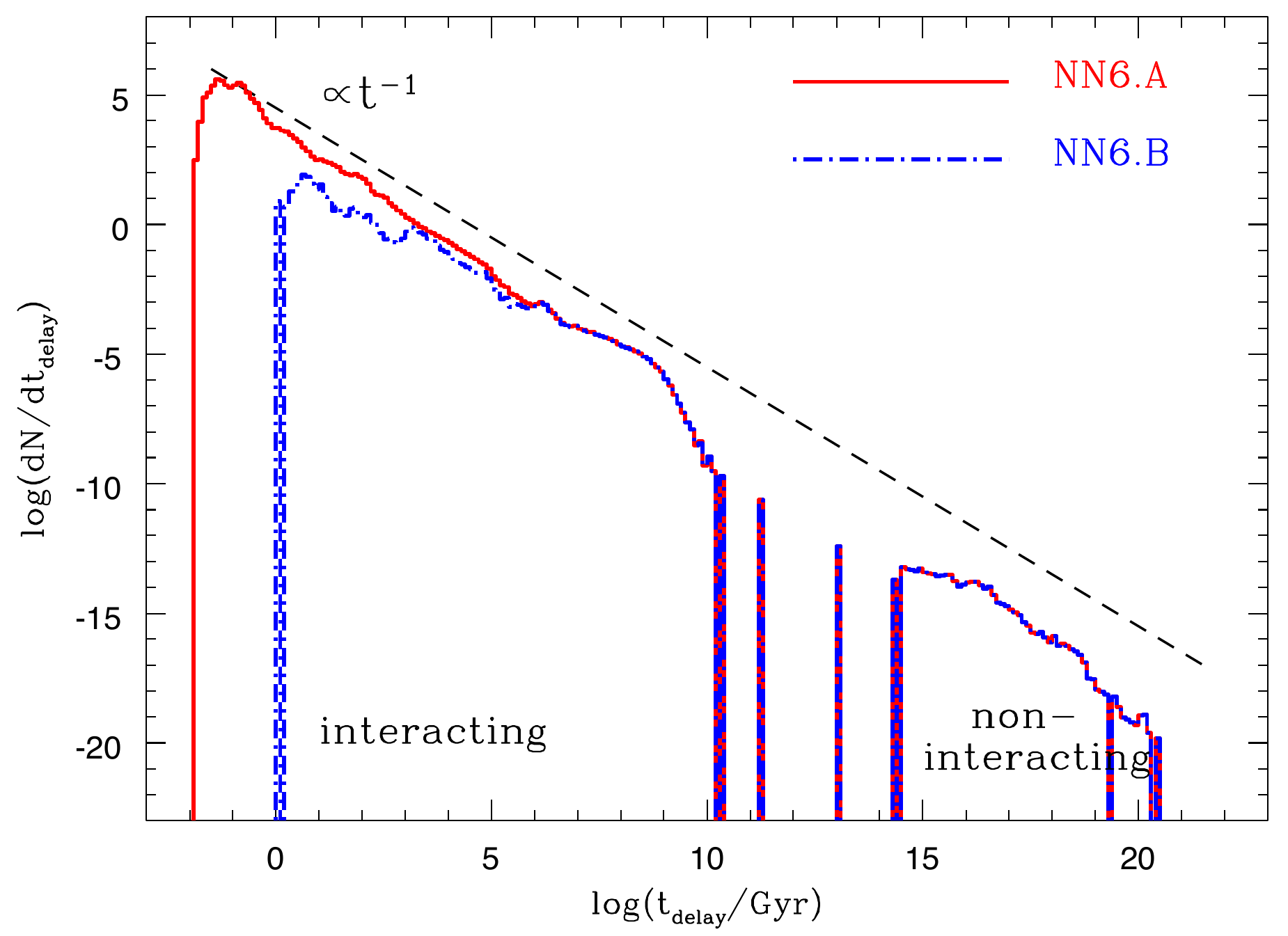}  
\caption{    
Model NN6: NS-NS delay time distribution. Labels same as in 
Figure~\ref{fig.tm10}. 
}
\label{fig.tnn6}
\end{figure}

Figures~\ref{fig.tnn13}, ~\ref{fig.tnn1}, ~\ref{fig.tnn6} show delay time 
distributions for NS-NS binaries in models NN13, NN1, NN6. These models include 
very low natal kicks as imposed by ~\cite{Bray2018} formula (NN1, NN6) or zero 
natal kicks (NN13) for all CC and ECS NSs. Additionally, very high CE efficiency 
is assumed ($\alpha_{\rm CE}=10$) in models NN13 and and NN6. These are rather 
extreme assumptions, as some NSs in NS-NS binaries are believed to receive at 
least small to moderate natal kicks, and it is rather unlikely that there is
as much extra energy in binary as $10$ times orbital energy available for CE 
ejection. Ignoring this for the moment we examine corresponding delay time
distributions. Although these distributions also follow power-law trend 
($t_{\rm delay}^{-1}$) there are some differences from other models. Notably
the distributions show some bumps, some of which peak around $10$ Gyr, and
distributions for submodels B start at rather late times ($\sim 1$ Gyr)
reaching peak at $10$ Gyr. This can possibly help to deliver high NS-NS
merger rates at late times after star formation. In fact, for model NN13.A
and NN6.A the current NS-NS merger rate from elliptical galaxies within
LIGO/Virgo reach is $\sim 156$--$561\gpy$, and thus consistent with
LIGO/Virgo estimate. Even model NN1.B rate from elliptical galaxies
($50.6\gpy$) is only factor of $2$ below LIGO/Virgo $90\%$ confidence level
lower limit. Obviously small NS natal kicks do not disrupt progenitor 
binaries increasing merger rates, while high CE efficiency allows for the 
increased formation of close NS-NS binaries with long delay times.

\subsection{Current NS-NS Galactic merger time distribution}

Note that the current NS-NS Galactic merger time distribution (observable) 
is different from merger time distribution at NS-NS formation (intrinsic). 
We transform our model NS-NS delay time distributions through continuous 
Galactic disk star formation applying radio-detectability criteria to obtain 
the current NS-NS Galactic merger time distributions (see Sec.~\ref{sec.crit2}).
 
\begin{figure}   
\hspace*{-0.3cm}   
\includegraphics[width=9.2cm]{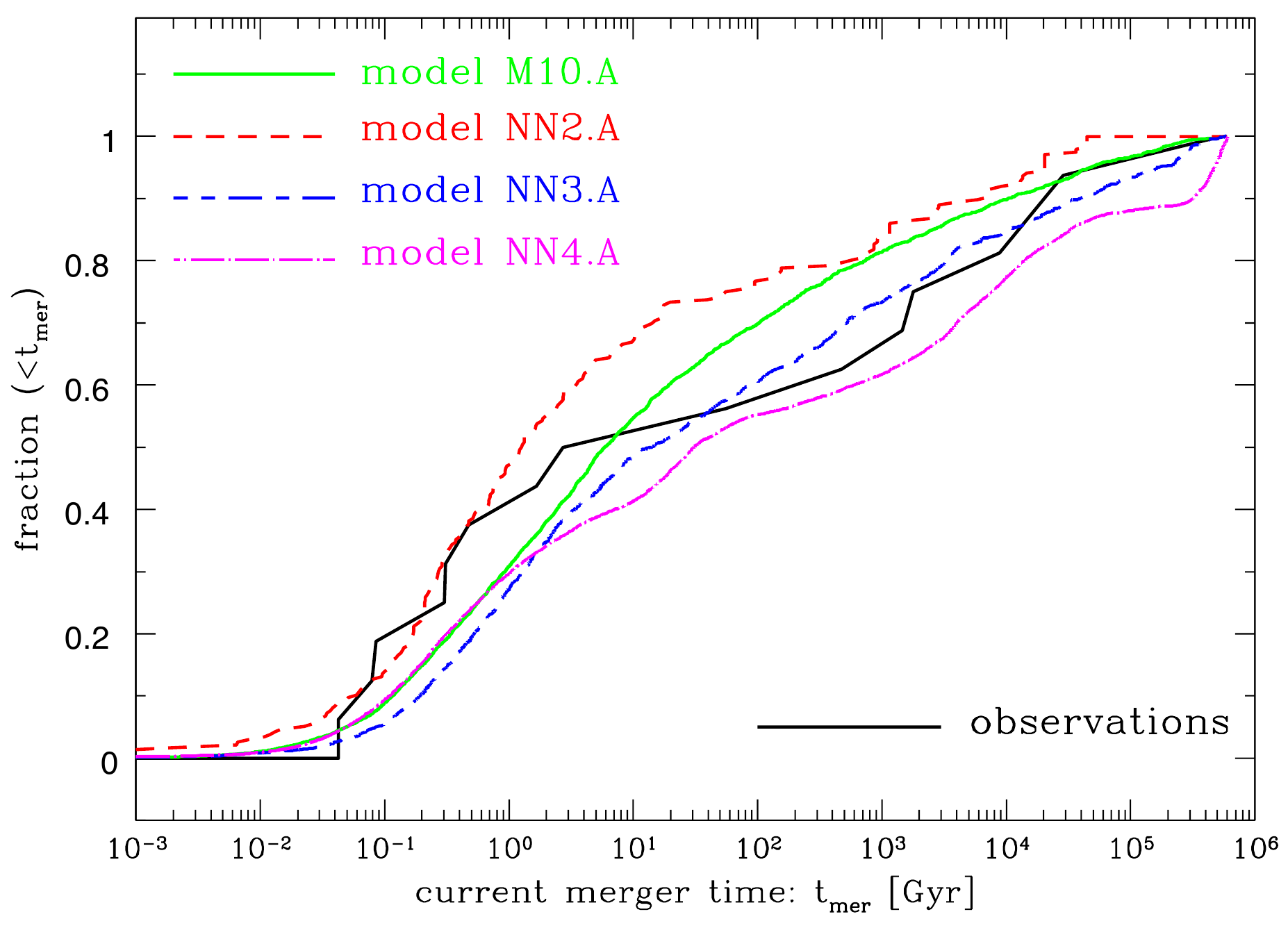}  
\caption{    
The cumulative distribution of the observed current merger times of 
NS-NS systems in the Milky Way and the ones predicted by models
M10.A, NN2.A, NN3.A and NN4.A with the inclusion of selection effects 
described in Section~\ref{sec.crit}. 
}
\label{fig.m10A}
\end{figure}

\begin{figure}   
\hspace*{-0.3cm}   
\includegraphics[width=9.2cm]{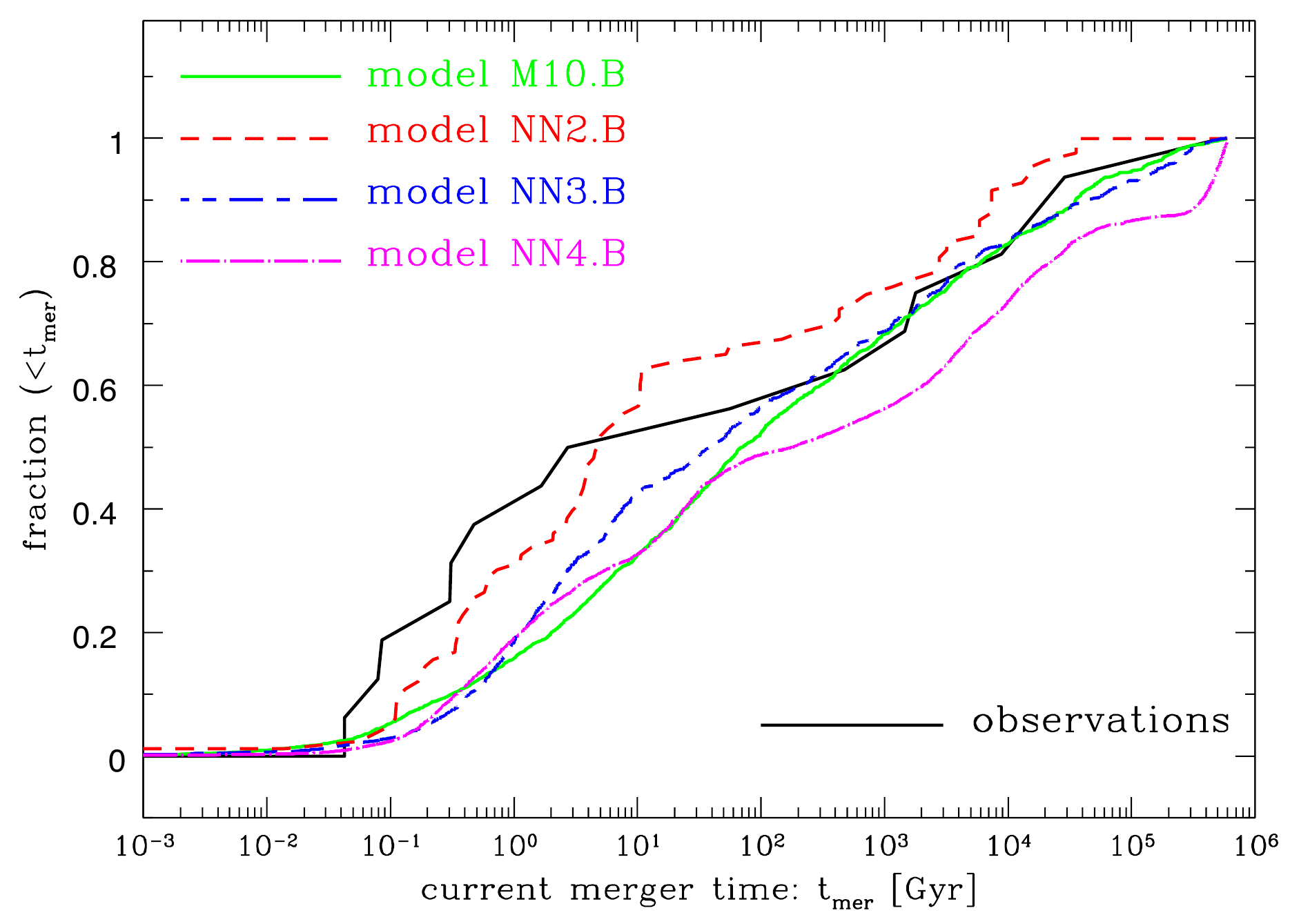}  
\caption{Same as Figure~\ref{fig.m10A} but for models M10.B, NN2.B, NN3.B
and NN4.B.}
\label{fig.m10B}
\end{figure}

Figures~\ref{fig.m10A} and ~\ref{fig.m10B} show the cumulative distribution of 
current Galactic NS-NS merger times for models M10, NN2, NN3, and NN4
for both submodels A and B. For comparison we also show the observed cumulative 
distribution of the $16$ Galactic field NS-NS binaries (see Tab.~\ref{tab.obs}). 
These models encompass rather broad assumptions on natal kicks; from all NSs
receiving high natal kicks (NN2; $\sigma=265\kms$ with no decrease due to
fallback), through high CC NS natal kicks and no ECS NS natal kicks (NN3), to 
no natal kicks at all (NN4). These models also probe the conservativeness of 
RLOF from $80\%$ mass loss (NN2, NN3, NN4) to $50\%$ (M10). We note that these 
model distributions are very similar to the observed distribution. In particular, 
all these model distributions have a significant probability ($\sim15\%$--$87\%$) 
of being drawn from the same underlying distribution as the observed sample: 
see KS test p-values given in Table~\ref{tab.stat10}. 

In Table~\ref{tab.stat10} we also list the fraction of short merger time NS-NS 
systems versus long merger time systems. We choose $30$ Gyr as a dividing line 
between short and long merger time systems. This time corresponds to the 
mid-point between short and long merger times systems known in the Galaxy. We 
find that many models are close to the $50\%$--$50\%$ observed ratio of 
short--long merger time systems (see Tab.~\ref{tab.obs}). Specifically, the 
models discussed above (M10, NN2, NN3, NN4) show fraction of long merger time 
systems in the range: $27$--$58\%$.   

\begin{figure}   
\hspace*{-0.3cm}   
\includegraphics[width=9.2cm]{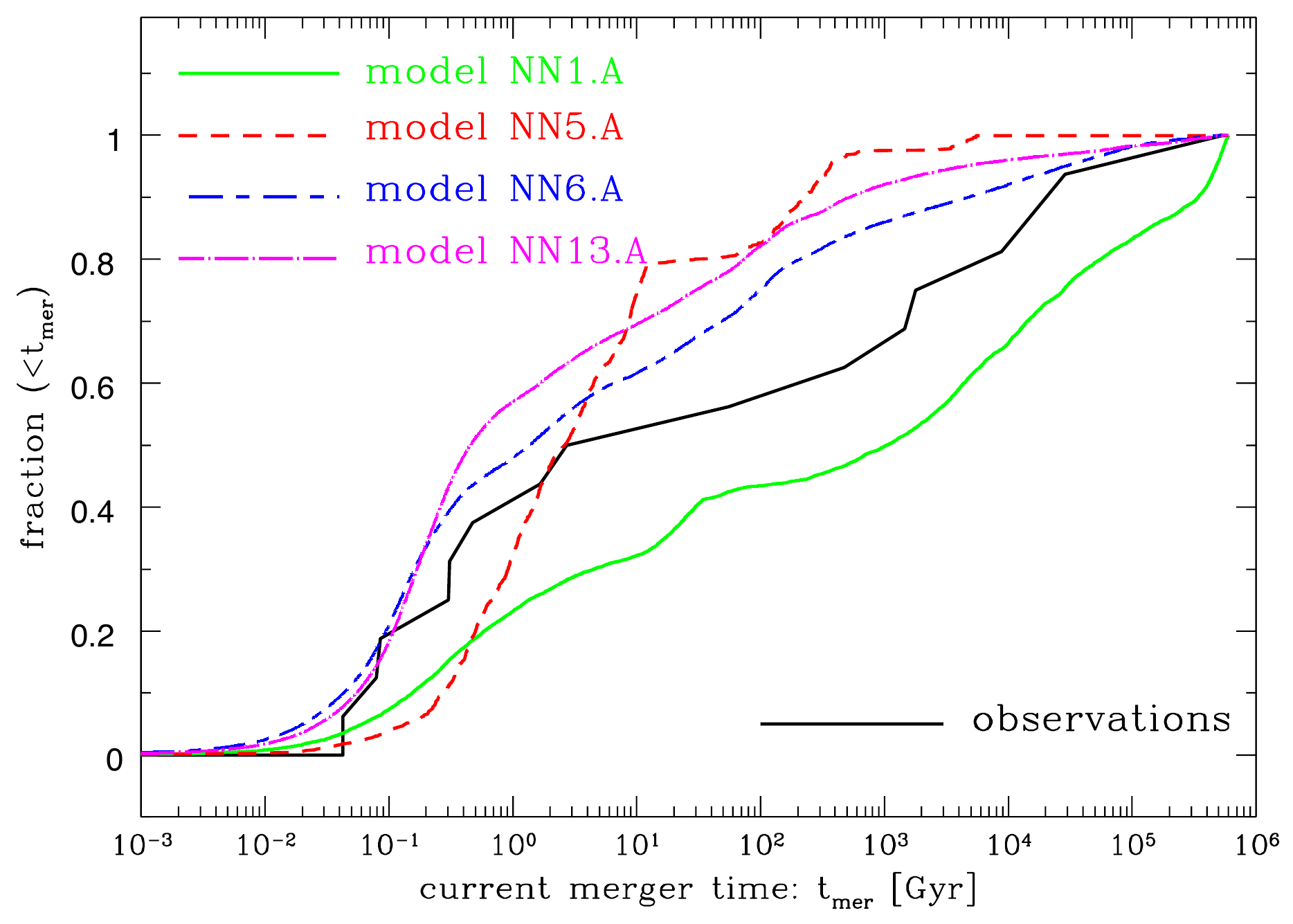}  
\caption{Same as Figure~\ref{fig.m10A} but for models NN1.A, NN5.A, NN6.A 
and NN13.A.}
\label{fig.m11A}
\end{figure}

\begin{figure}   
\hspace*{-0.3cm}   
\includegraphics[width=9.2cm]{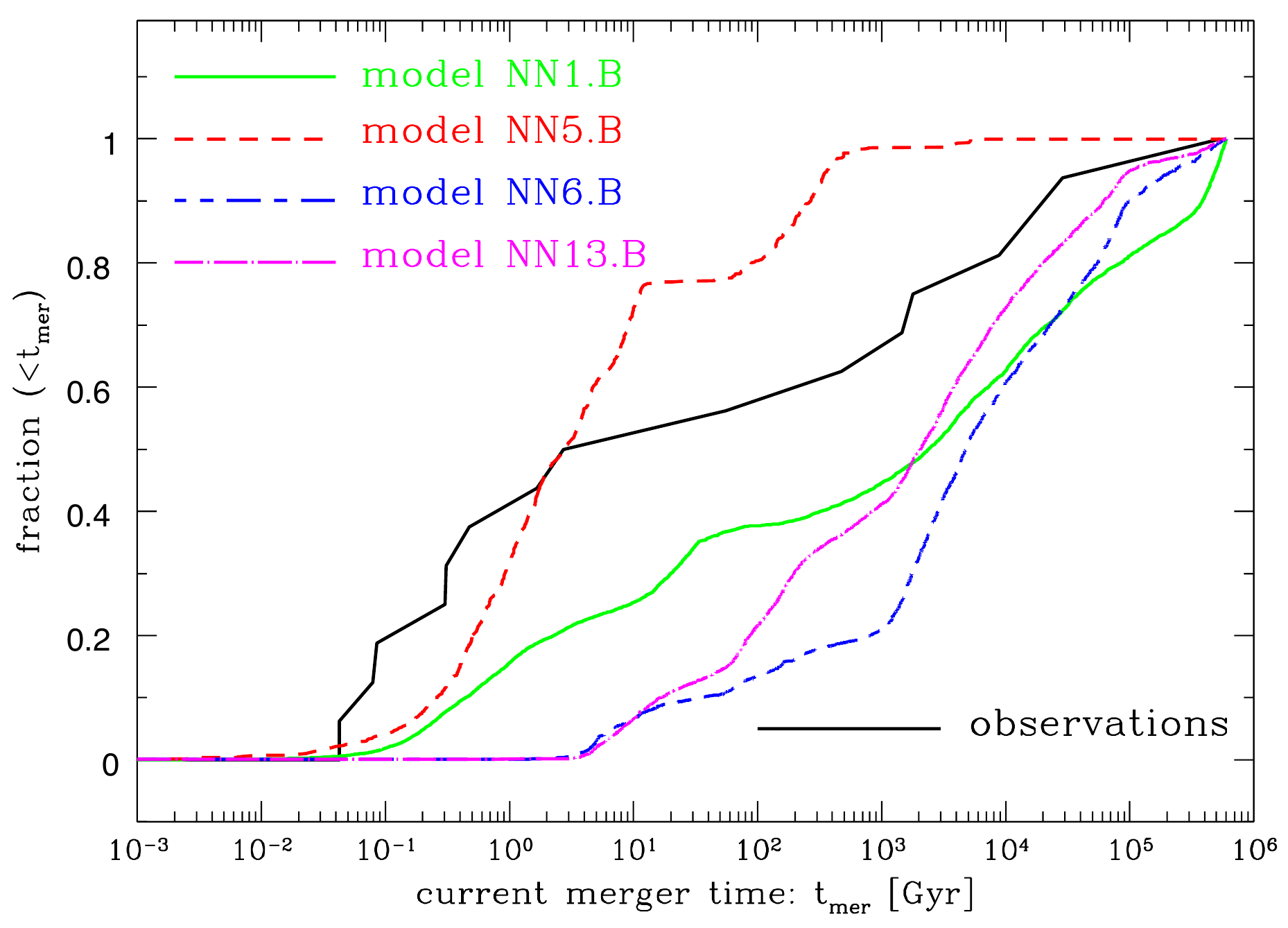}  
\caption{Same as Figure~\ref{fig.m10A} but for models NN1.B, NN5.B, NN6.B
and NN13.B.}
\label{fig.m11B}
\end{figure}

Figures~\ref{fig.m11A}, and ~\ref{fig.m11B} show the cumulative distribution 
of current Galactic NS-NS merger times for models NN1, NN5,NN6, and NN13. 
These models encompass some of our extreme assumptions on input physics. In
models NN1, NN5, NN6 we employ ~\cite{Bray2018} natal kicks while varying CE 
efficiency: $\alpha_{\rm CE}=0.1,1.0,10$, respectively. In model NN13 NSs do not 
receive any natal kicks and very high CE efficiency is used  $\alpha_{\rm CE}=10$.
We note that these models do not match observations as well as the other models 
(M10, NN2, NN3, and NN4 shown in Fig.~\ref{fig.m10A} and ~\ref{fig.m10B}). 
In particular, the probability that these models are drawn from the same
distribution as the observed sample can be very small and is found in the 
range $\sim 0.01$--$37\%$ for all these four models. The fraction of the long 
current merger time systems varies in wide range for these models $20$--$90\%$ 
for these models (see Tab.~\ref{tab.obs}).

\subsection{Milky Way NS-NS merger rate vs LIGO/Virgo rate}

In Table~\ref{tab.rates} we report the Galactic merger rate of NS-NS systems
(see Sec.~\ref{sec.crit1}) along with the corresponding rate density on 
NS-NS mergers from elliptical galaxies in local Universe (see Sec.
~\ref{sec.crit3}) for all our models.
The range reported in Galactic model NS-NS merger rates corresponds to an 
assumption on the star forming mass in the Galactic disk. The low rates 
correspond to the disk mass of $3.5 \times 10^{10} \msun$ that was used in 
our previous estimates \citep{Belczynski2010a,Dominik2012} while the high 
rates correspond to the disk mass of $5.17 \times 10^{10} \msun$ recently 
estimated by \cite{Licquia2015}.
The range reported in model NS-NS merger rate density in elliptical galaxies 
corresponds to assumed age of ellipticals (their stellar populations): all 
ellipticals 10 Gyr old (left hand side values) or age distributed uniformly 
in range $1$--$11$ Gyr (right hand side values). 

For comparison with models we also list empirical estimates in Table~\ref{tab.rates}. 
The empirical Galactic merger rate estimates are based on the $8$ Galactic 
NS-NS systems with merger times shorter than the Hubble time: $28$--$72\Mpy$
~\citep{Pol2018} and $6.6$--$190\Mpy$ (see App.~\ref{sec.nsnsrate}). 
For empirical NS-NS merger rate density in elliptical galaxies in local 
Universe we adopt LIGO/Virgo estimate: $110$--$3840\gpy$. 

Investigation of Table~\ref{tab.rates} allows us to note:
\begin{enumerate}

\item Some models may be rejected based on comparison with Galactic NS-NS 
merger rates: NN2.B (high \cite{Hobbs2005} natal kicks for all NSs with HG
CE not allowed), NN14.B (high \cite{Hobbs2005} natal kicks with fallback for 
all NSs with HG CE not allowed), NN7.B (medium \cite{Hobbs2005} natal kicks
for all NSs with HG CE not allowed), NN4.A (no natal kicks for all NSs with 
HG CE allowed), NN13.A (no natal kicks for all NSs and very high CE efficiency 
with HG CE allowed), NN1.A (\cite{Bray2018} natal kicks with HG CE allowed), 
and NN6.A/B (\cite{Bray2018} natal kicks with very high CE efficiency 
independent of treatment of CE survival with HG donor).

\item Two models generate NS-NS merger rate densities in local Universe 
elliptical galaxies high enough to overlap with LIGO/Virgo estimate: NN13.A 
(zero natal kicks for all NSs with HG CE allowed) and NN6.A (\cite{Bray2018} 
natal kicks with very high CE efficiency and HG CE allowed). 
However, these models produce extremely high Galactic NS-NS merger rates 
($\gtrsim 1000\Mpy$) that significantly exceed empirical Milky Way rate 
estimates ($\lesssim 200\Mpy$).

\item Majority of models generate NS-NS merger rates that are consistent with 
the Milky Way empirical merger rate estimates: NN2.A, NN14.A, NN7.A, NN3.A/B,  
NN8.A/B, M10.A/B, NN11.A/B, NN9.A/B, NN10.A/B, NN12.A/B, NN4.B, NN13.B, NN5.A/B, 
and NN1.B. None of these models deliver rate of NS-NS mergers in elliptical 
galaxies consistent with LIGO/Virgo estimate. 
Note that for comparisons we use larger Galactic NS-NS merger model rates as
they correspond to updated mass of the Galactic disk. 

\end{enumerate}

\section{Conclusions}
\label{sec.5}

We have analyzed the validity of evolutionary predictions in the context of
the recent gravitational-wave detection of the first NS-NS merger and taking
into account the already rich population of $16$ Galactic field NS-NS binaries. 
Generally, evolutionary predictions are consistent with the Galactic population 
of NS-NS binaries recovering the Galactic merger rates and observed merger time 
distribution. However, at the same time, the models that are in agreement with 
the Galactic observations, generate local cosmic NS-NS merger rates in 
elliptical galaxies that are not consistent with LIGO/Virgo estimate. 

Evolutionary models predict that currently (low redshifts) the majority of NS-NS
mergers should be found in systems with ongoing or recent star formation, due to 
a steep-power law ($\propto t^{-1}$) delay time distribution
~\citep[see also][]{Richard2010}. Note that Advanced LIGO/Virgo even at its 
design sensitivity will be only able to discover NS-NS mergers at low-redshifts 
($z\lesssim0.1$). 
For higher redshifts, models predict the increasing contribution of elliptical 
hosts, as the time between star formation in elliptical galaxies ($\sim 1$--$10$
Gyr ago) and a given redshift decreases. Our models appear to be broadly 
consistent with observations of short GRBs~\citep{Richard2008a}. \cite{Fong2013} 
estimated the frequency of short GRBs among late type hosts ($60$--$80\%$) and 
early type hosts galaxies ($20$--$40\%$). 
This analysis includes also host-less and inconclusive short GRB cases. If only 
confirmed-host short GRBs are used, the fraction of short GRBs in early type 
hosts (e.g., ellipticals) drops down to $\sim 20\%$.

There is a set of population synthesis calculations performed with the BPASS 
code~\citep{Eldridge2017} that generates large Galactic NS-NS merger rates: 
a most likely value of $386\Mpy$ with uncertainty range of $149$--$543\Mpy$
~\citep{Bray2018,Eldridge2018}. This estimate was obtained for solar metallicity 
($Z=0.02$) and a Galactic disk mass of $3.5 \times 10^{10}\msun$. If we correct 
this rate for more realistic Galactic disk mass: $5.17 \times 10^{10}\msun$
~\citep{Licquia2015}, then the BPASS NS-NS merger rate is $571\Mpy$ with an 
uncertainty range: $221$--$804\Mpy$. This high rate seem to be at odds with 
the empirical Galactic merger rates: $28$--$72\Mpy$ \citep{Pol2018} and 
$6.6$--$190\Mpy$ (see App.~\ref{sec.nsnsrate}). 
This high rate is obtained with a new prescription of NS natal kicks 
that is based on the ratio of supernova ejecta mass to NS mass predicted 
in models. We have tested this new prescription in our calculations. And 
in fact this prescription results in almost no NS natal kicks increasing
significantly NS-NS merger rate.   

We find that it is possible to construct evolutionary models that generate
NS-NS merger rates in elliptical galaxies that are consistent with the 
LIGO/Virgo estimate. These models employ no or very low natal NS kicks
(e.g., ~\cite{Bray2018} natal kicks) and very high CE efficiency. However, 
these models are not consistent with the Galactic population of NS-NS 
binaries. For example, Galactic merger rates generated in these models are 
much higher ($\sim 1000\Mpy$) than estimated from observations ($<200\Mpy$).

Thus we are left with the tension between evolutionary models supported by the 
Galactic NS-NS observations and the LIGO/Virgo early detection of a NS-NS merger 
in an old host galaxy. Potential solutions of this problem may include: 

\begin{enumerate}

\item The LIGO/Virgo detection was a statistical fluctuation, and the following
detections will be associated with regions with ongoing or recent star
formation. After all, models do predict NS-NS mergers in old hosts, alas at
low rates. If this is the case, this will be quickly resolved by near-future
observations by LIGO/Virgo at increased sensitivity during the O3 run in 2019. 

\item If the above is not true then it it possible that the solution sits
in a part of multi-dimensional parameter space that we did not probe with our
very limited models. Massive multi-dimensional parameter studies
~\citep[e.g.,][]{Richard2008b} are needed to confirm or reject such hypothesis. 
Here we have only attempted to probe a small part of parameter space to serve 
as an initial step for such future studies. 

\item If the above is not true then if LIGO/Virgo keep detecting NS-NS 
mergers in old hosts (as it may be indicated by GRB 150101B; see Sec.~\ref{sec.1}) 
then this will call for a revision of models of isolated binary evolution. 
Such a revision would need to preserve typically short delay times produced 
by current models for star forming galaxies (e.g., the Milky Way), but it 
will need to generate typically long delay times for old galaxies (e.g., 
ellipticals). 

\item If the above is not true then isolated binary evolution model that 
connect NS-NS formation directly to star forming mass is not the correct 
solution in the case of GW170817. It was already proposed that dynamical 
interactions between stars (or compact objects) in two merging galaxies may 
induce enhanced NS-NS formation~\citep{Palmese2017}. In this context, even 
with no significant star formation in a conceivable recent minor merger in 
NGC 4993, it may have been possible that a NS-NS  merger formed through 
enhanced dynamical interactions. 

\end{enumerate}

\begin{table}
\caption{Properties of NS-NS binaries\tablefootmark{a}}
\centering
\begin{tabular}{c| c c c c}
\hline\hline
Name         & short\tablefootmark{b} &   long & p-value\tablefootmark{c} & $\cal{R}_{\rm MW}$ [Myr$^{-1}$] \\
\hline\hline
observations &                 $50\%$ & $50\%$ &       & $28$--$72$\tablefootmark{d}   \\
             &                        &        &       & $6.6$--$190$\tablefootmark{e} \\
\hline

NN2.A        &                 $73\%$ & $27\%$ & 0.260 & 13.5--20.0  \\
NN2.B        &                 $64\%$ & $36\%$ & 0.813 &   0.9--1.3  \\
             &                        &        &       &             \\ 
NN14.A       &                 $82\%$ & $18\%$ & 0.021 & 22.6--33.4  \\
NN14.B       &                 $77\%$ & $23\%$ & 0.067 &   1.5--2.2  \\
             &                        &        &       &             \\ 
NN7.A        &                 $62\%$ & $38\%$ & 0.445 & 32.4--48.0  \\
NN7.B        &                 $53\%$ & $47\%$ & 0.532 &   3.1--4.6  \\
             &                        &        &       &             \\ 
NN3.A        &                 $54\%$ & $46\%$ & 0.587 & 38.4--56.8  \\
NN3.B        &                 $48\%$ & $52\%$ & 0.165 & 10.8--16.0  \\
             &                        &        &       &             \\ 
NN8.A        &                 $43\%$ & $57\%$ & 0.329 & 45.0--66.6  \\
NN8.B        &                 $34\%$ & $66\%$ & 0.091 & 10.6--15.7  \\
             &                        &        &       &             \\ 
M10.A        &                 $63\%$ & $37\%$ & 0.367 & 53.6--79.3  \\
M10.B        &                 $42\%$ & $58\%$ & 0.145 & 17.4--25.8  \\
             &                        &        &       &             \\ 
NN11.A       &                 $57\%$ & $43\%$ & 0.731 & 61.1--90.4  \\
NN11.B       &                 $52\%$ & $48\%$ & 0.376 &  7.8--11.5  \\
             &                        &        &       &             \\ 
NN9.A        &                 $51\%$ & $49\%$ & 0.875 &  67.6--100  \\
NN9.B        &                 $46\%$ & $54\%$ & 0.318 & 11.0--16.3  \\
             &                        &        &       &             \\ 
NN10.A       &                 $36\%$ & $64\%$ & 0.246 &  76.9--114  \\
NN10.B       &                 $29\%$ & $71\%$ & 0.058 & 16.0--23.7  \\
             &                        &        &       &             \\ 
NN12.A       &                 $50\%$ & $50\%$ & 0.967 &   126--186  \\
NN12.B       &                 $43\%$ & $57\%$ & 0.283 & 21.8--32.3  \\
             &                        &        &       &             \\ 
NN4.A        &                 $49\%$ & $51\%$ & 0.873 &   251--371  \\
NN4.B        &                 $43\%$ & $57\%$ & 0.238 & 48.9--72.4  \\
             &                        &        &       &             \\ 
NN13.A       &                 $75\%$ & $25\%$ & 0.044 & 1208--1788  \\
NN13.B       &                 $13\%$ & $87\%$ & $\sim10^{-4}$ &    6.7-9.9  \\
             &                        &        &       &             \\ 
\hline
NN5.A        &                 $80\%$ & $20\%$ & 0.008 &  11.9--17.6 \\
NN5.B        &                 $77\%$ & $23\%$ & 0.007 &  11.5--17.0 \\
             &                        &        &       &             \\ 
NN1.A        &                 $40\%$ & $60\%$ & 0.368 &   179--265  \\
NN1.B        &                 $34\%$ & $66\%$ & 0.108 & 37.0--54.8  \\
             &                        &        &       &             \\ 
NN6.A        &                 $67\%$ & $33\%$ & 0.156 &  961--1422  \\
NN6.B        &                 $10\%$ & $90\%$ & $\sim10^{-4}$ &   4.1--6.1  \\
\hline
\hline
\end{tabular}
\label{tab.stat10}
\tablefoot{ \\
\tablefootmark{a}{Comparison of observed and model current merger time distributions.}\\
\tablefootmark{b}{short: $t_{\rm mer}<30$ Gyr, long: $t_{\rm mer}>30$ Gyr.}\\
\tablefootmark{c}{Probability that observations and model were drawn from
the same delay time distribution (KS test).}\\
\tablefootmark{d}{Estimate from \cite{Pol2018} with peak probability value 
of $\cal{R}_{\rm MW}=$42 Myr$^{-1}$.}\\
\tablefootmark{e}{Estimate presented in App.~\ref{sec.nsnsrate}.}\\
}
\end{table}

\begin{table*}
\caption{NS-NS merger rates: sorted by natal kick and increasing Galactic rates}
\centering
\begin{tabular}{c| c c c c | c | c}
\hline\hline
Name & CC kick\tablefootmark{a} & ECS kick\tablefootmark{b} & $\alpha_{\rm CE}$\tablefootmark{c} & 
(acc/eje)$_{\rm RLOF}$\tablefootmark{d} & $\cal{R}_{\rm MW}$ [Myr$^{-1}$]\tablefootmark{e} & 
$\cal{R}_{\rm ell}$ [$\gpy$]\tablefootmark{f} \\
\hline\hline
observations &                      &              &      &         & $28$--$72$\tablefootmark{g}  & $110$--$3840$\tablefootmark{i} \\
             &                      &              &      &         & $6.6$--$190$\tablefootmark{h} & \\
\hline
             &                      &              &      &         &                  & \\
NN2.A        & Hobbs: $265\kms$     & OFF: --      &  1.0 & 0.2/0.8 & $13.5$--$20.0$   & $0.8$--$2.3$ \\
NN2.B        & Hobbs: $265\kms$     & OFF: --      &  1.0 & 0.2/0.8 & $0.9$--$1.3$     & $0.8$--$2.3$ \\
             &                      &              &      &         &                  & \\
NN14.A       & HobbsFB: $265\kms$   & OFF: --      &  1.0 & 0.2/0.8 & $22.6$--$33.4$   & $0.8$--$3.0$ \\
NN14.B       & HobbsFB: $265\kms$   & OFF: --      &  1.0 & 0.2/0.8 & $1.5$--$2.2$     & $0.8$--$2.0$ \\
             &                      &              &      &         &                  &        \\
NN7.A        & Hobbs: $133\kms$     & ON: $66\kms$ &  1.0 & 0.2/0.8 & $32.4$--$48.0$   & $1.2$--$6.2$ \\
NN7.B        & Hobbs: $133\kms$     & ON: $66\kms$ &  1.0 & 0.2/0.8 & $3.1$--$4.6$     & $1.2$--$4.1$ \\
             &                      &              &      &         &                  & \\
NN3.A        & HobbsFB: $265\kms$   & ON: $0\kms$  &  1.0 & 0.2/0.8 & $38.4$--$56.8$   & $6.3$--$21.0$ \\
NN3.B        & HobbsFB: $265\kms$   & ON: $0\kms$  &  1.0 & 0.2/0.8 & $10.8$--$16.0$   & $5.9$--$18.9$ \\
             &                      &              &      &         &                  &        \\
NN8.A        & Hobbs: $133\kms$     & ON: $0\kms$  &  1.0 & 0.2/0.8 & $45.0$--$66.6$   & $8.3$--$19.6$ \\
NN8.B        & Hobbs: $133\kms$     & ON: $0\kms$  &  1.0 & 0.2/0.8 & $10.6$--$15.7$   & $7.5$--$15.6$ \\
             &                      &              &      &         &                  & \\
M10.A        & HobbsFB: $265\kms$   & ON: $0\kms$  &  1.0 & 0.5/0.5 & $53.6$--$79.3$   & $11.4$--$51.4$ \\
M10.B        & HobbsFB: $265\kms$   & ON: $0\kms$  &  1.0 & 0.5/0.5 & $17.4$--$25.8$   & $18.5$--$22.1$ \\
             &                      &              &      &         &                  &        \\
NN11.A       & Hobbs: $66\kms$      & OFF: --      &  1.0 & 0.2/0.8 & $61.1$--$90.4$   & $4.7$--$13.1$ \\
NN11.B       & Hobbs: $66\kms$      & OFF: --      &  1.0 & 0.2/0.8 & $7.8$--$11.5$    & $4.3$--$11.8$ \\
             &                      &              &      &         &                  & \\
NN9.A        & Hobbs: $66\kms$      & ON: $33\kms$ &  1.0 & 0.2/0.8 & $67.6$--$100$    & $3.9$--$18.4$ \\
NN9.B        & Hobbs: $66\kms$      & ON: $33\kms$ &  1.0 & 0.2/0.8 & $11.0$--$16.3$   & $3.9$--$16.3$ \\
             &                      &              &      &         &                  & \\
NN10.A       & Hobbs: $66\kms$      & ON: $0\kms$  &  1.0 & 0.2/0.8 & $76.9$--$114$    & $7.9$--$29.9$ \\
NN10.B       & Hobbs: $66\kms$      & ON: $0\kms$  &  1.0 & 0.2/0.8 & $16.0$--$23.7$   & $7.5$--$27.7$ \\
             &                      &              &      &         &                  & \\
NN12.A       & Hobbs: $33\kms$      & OFF: --      &  1.0 & 0.2/0.8 & $126$--$186$     & $13.4$--$33.1$ \\
NN12.B       & Hobbs: $33\kms$      & OFF: --      &  1.0 & 0.2/0.8 & $21.8$--$32.3$   & $13.4$--$31.5$ \\
             &                      &              &      &         &                  & \\
NN4.A        & Hobbs: $0\kms$       & OFF: --      &  1.0 & 0.2/0.8 & $251$--$371$     & $23.2$--$72.1$ \\
NN4.B        & Hobbs: $0\kms$       & OFF: --      &  1.0 & 0.2/0.8 & $48.9$--$72.4$   & $23.2$--$70.8$ \\
             &                      &              &      &         &                  & \\
NN13.A       & Hobbs: $0\kms$       & OFF: --      &   10 & 0.2/0.8 & $1208$--$1788$   & $186$--$561$ \\
NN13.B       & Hobbs: $0\kms$       & OFF: --      &   10 & 0.2/0.8 & $6.7$--$9.9$     & $29.9$--$25.2$ \\
\hline
             &                      &              &      &         &                  & \\
NN5.A        & BE18: $100/-170\kms$ & OFF: --      &  0.1 & 0.2/0.8 & $11.9$--$17.6$   & $11.8$--$22.9$ \\
NN5.B        & BE18: $100/-170\kms$ & OFF: --      &  0.1 & 0.2/0.8 & $11.5$--$17.0$   & $11.8$--$22.9$ \\
             &                      &              &      &         &                  & \\
NN1.A        & BE18: $100/-170\kms$ & OFF: --      &  1.0 & 0.2/0.8 & $179$--$265$     & $15.3$--$51.2$ \\
NN1.B        & BE18: $100/-170\kms$ & OFF: --      &  1.0 & 0.2/0.8 & $37.0$--$54.8$   & $15.3$--$50.6$ \\
             &                      &              &      &         &                  &        \\
NN6.A        & BE18: $100/-170\kms$ & OFF: --      &   10 & 0.2/0.8 & $961$--$1422$    & $156$--$471$ \\
NN6.B        & BE18: $100/-170\kms$ & OFF: --      &   10 & 0.2/0.8 & $4.1$--$6.1$     & $12.6$--$15.1$ \\
\hline
\hline
\end{tabular}
\label{tab.rates}
\tablefoot{ \\
\tablefootmark{a}{Core collapse SN NS natal kicks. 
Hobbs: Maxwellian distribution with a given 1-D $\sigma$; 
HobbsFB: Maxwellian distribution with a given 1-D $\sigma$ lowered by fall-back; 
BE18: Bray \& Eldridge kicks with a given $\alpha$ and $\beta$.}\\
\tablefootmark{b}{Electron capture SN NS natal kicks. 
OFF: all NSs form through CC SNe; 
ON: ECS allowed with a kick from Maxwellian distribution with a given 1-D
$\sigma$ with fall back as for CC kicks.}\\
\tablefootmark{c}{Common envelope efficiency}\\
\tablefootmark{d}{Mass fraction of donor mass transfer accreted by donor/ejected from 
binary during stable RLOF}\\
\tablefootmark{e}{The Milky Way NS-NS merger rate. Left--right hand side values 
correspond to Galactic disk mass of $3.5\times10^{10}\msun$--$5.17\times10^{10}\msun$.}\\
\tablefootmark{f}{Local cosmic ($z=0$) merger rate density for NS-NS systems formed 
only in elliptical galaxies. Left--right hand side values correspond to assumed age
of ellipticals (their stellar populations): all ellipticals 10 Gyr old -- age 
distributed uniformly in range $1$--$11$ Gyr.}\\
\tablefootmark{g}{$90\%$ confidence level, with peak value of $42\Mpy$ from
\cite{Pol2018}.}\\
\tablefootmark{h}{Conservative rate range described in App.~\ref{sec.nsnsrate}.}\\
\tablefootmark{i}{LIGO/Virgo $90\%$ confidence level, with peak value of $\sim 1000\gpy$}\\
}
\end{table*}

\begin{acknowledgements}

KB acknowledges support from the Polish National Science Center (NCN) grants
OPUS (2015/19/B/ST9/01099), Maestro (2015/18/A/ST9/00746) and LOFT/eXTP 
(2013/10/M/ST9/00729).
We would like to thank thousands of {\tt Universe@home} users that have provided 
their personal computers and phones for our simulations, and in particular to K. 
Piszczek (program IT manager). TB acknowledges support from the 
grant FNP grant TEAM-2016-3/19.

\end{acknowledgements}

\appendix

\section{NS-NS rate estimate}
\label{sec.nsnsrate}

We estimate the galactic NS-NS merger rate $\cal{R}_{\rm MW}$ from the observed 
number of coalescing NS-NS binaries ($N_{psr}=8$) using two simplifying and 
phenomenologically motivated assumptions. First, we assume that the NS-NS 
population can be characterized by a single population-averaged probability $S$ 
to be detected in radio, expressed as $S =\langle (f_b N_{psr})^{-1} \rangle$ 
where $1/N_{psr}$ is the probability of detecting such a pulsar if it were 
pointing towards us, and $1/f_b$ is the probability the pulsar beam is pointing 
towards us. From the observed sample and previous work, we for simplicity adopt 
a fiducial value of $S=2\times 10^{-3}$, though this is uncertain by as much as 
$50\%$. Second, motivated by past and present theoretical modeling, we assume 
that NS-NS merger time  distribution $dP/d\tau$ after a burst of star formation 
is proportional to $1/\tau$, corresponding to a cumulative fraction of mergers 
$P(<\tau)$ which is roughly linear in $\log \tau$. Using this delay time 
distribution, we can estimate the fraction of binaries born at time $t_b$ which 
merge on or after the present day, as 
$P_{alive}(t_b)=\int_{T}^{\infty} (dP/d\tau) d\tau = 1-P(<T-t_b)$, where $T$ is 
the age of the Milky Way disk. Using these two simplifying assumptions, we can 
show a steady-state merger rate $\cal{R}_{\rm MW}$ will lead on average to 
$\mu=\cal{R}_{\rm MW}$ $\int_0^T dt_b P_{ok}(t_b)\simeq 1.26{\rm Gyr}\times \cal{R}_{\rm MW}$  
NS-NS binaries overall present in the Milky Way disk at the present day, with 
$\mu_{\rm tight}=0.081 {\rm Gyr}\times \cal{R}_{\rm MW}$ NS-NS binaries expected 
to coalesce in the near future. Accounting for selection biases, we estimate the 
merger rate from  $N_{psr}$ observed Galactic coalescing NS-NS binaries via  
$\cal{R}_{\rm MW}$ $=N_{psr}/\mu S$.  
Using the two estimates $\mu,\mu_{\rm tight}$ to bracket our uncertainty, we 
arrive at a NS-NS merger rate estimate of 
$\cal{R}_{\rm MW}$ $=A(S/2\times 10^{-3}) \Mpy$ where $A$ is between $10$--$95$. 
Allowing for uncertainty in $S$ broadens this estimate by $50\%$, leading to a 
conservative range of NS-NS merger rate $\cal{R}_{\rm MW}$ $=6.6$--$190\Mpy$.

\section{Radio and recycled pulsar lifetimes}

To obtain the distribution of the radio lifetimes of radio pulsars and recycled pulsars 
in NS-NS systems, we use PsrPopPy~\citep{Bates2014} to simulate the observed pulsar 
population. In these simulations, we use the Parkes Multibeam survey which was able to 
detect $\sim 1100$ pulsars. We use this survey to model pulsar survey parameters 
\citep[see Sec.~4.1 in][]{Bates2014}.
            
\begin{figure}
\hspace*{-0.3cm}
\includegraphics[width=9.2cm]{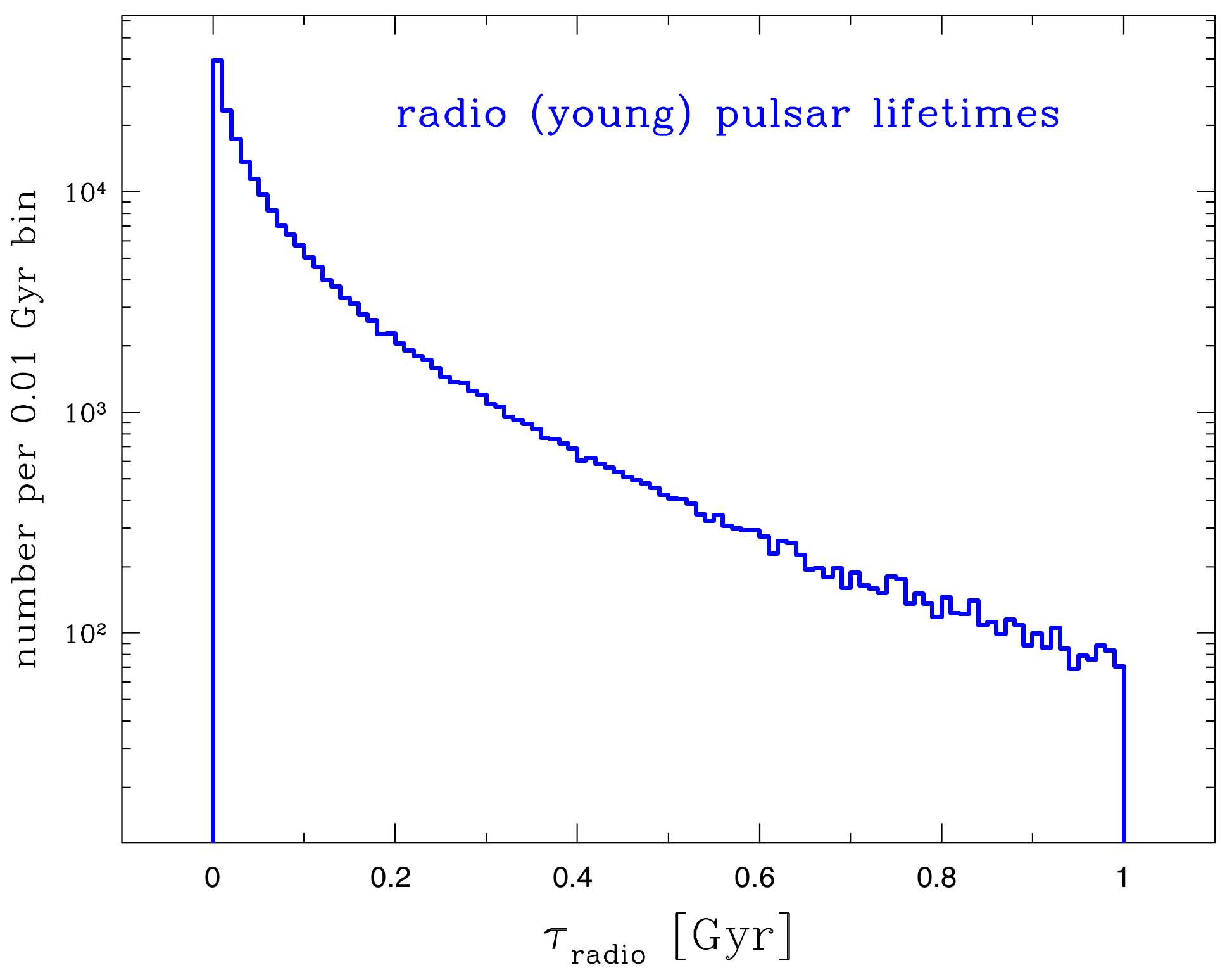}
\caption{Radio (young) pulsar lifetime distribution. Note that majority ($90\%$) of 
radio pulsars have short lifetimes ($t_{\rm radio}<315$ Myr). The average lifetime 
is $t_{\rm radio,ave}=116$ Myr, while median is $t_{\rm radio,med}=52.2$ Myr.}    
\label{fig.radio}
\end{figure}

To simulate the radio pulsar population, we use a normal distribution for the periods of 
the pulsars, with mean spin period $\left< P_{\rm s} \right> = 300$~ms and standard 
deviation $\sigma_{\rm P} = 100$~ms. We similarly assume a log-normal magnetic field 
distribution for these pulsars, with mean $\left< {\rm log}_{10}B {\rm (G)} \right> =12.65$ 
and standard deviation $\sigma_{{\rm log}_{10}B} = 0.55$.
These pulsars are assigned a lifetime, $\tau_{\rm r}$, which are drawn randomly from a 
uniform distribution between zero and the age of the Milky Way ($\sim$10~Gyr). The 
period and period derivative of the pulsars are evolved using the magnetic dipole model 
\citep[see Sec.~3 in][]{Bates2014} over their lifetime. If these pulsars cross the 
``death-line"~\citep{Chen1993} during their lifetime, they will not be observable and 
are removed from the simulation. Finally, we determine if the pulsars that are alive are 
detectable in the Parkes Multibeam survey~\citep[see Sec.~4 in][]{Bates2014}. We keep 
generating such pulsars in the simulation until we detect the $1059$ radio pulsars detected 
by the Parkes Multibeam Survey~\citep{Manchester2005}. The distribution of the radio lifetimes of 
the entire population so generated (i.e., not only the $1059$ pulsars detected in the survey) 
is shown in Fig.~\ref{fig.radio}.
        
\begin{figure}
\hspace*{-0.3cm}
\includegraphics[width=9.2cm]{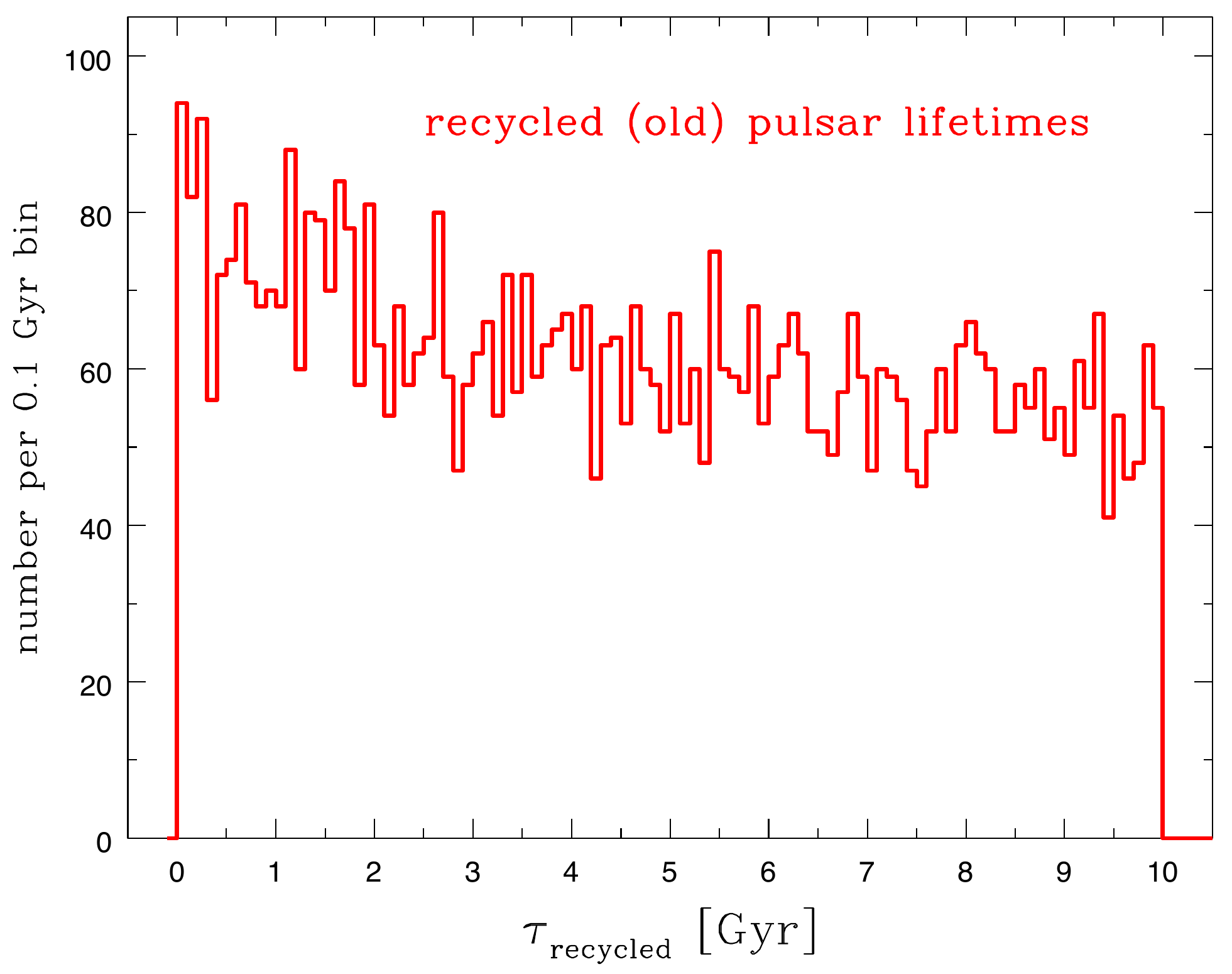}
\caption{Recycled (old) pulsar lifetime distribution. Note that the pulsar
lifetimes are distributed rather uniformly in a very broad range 
($t_{\rm recycled}=0$--$10$ Gyr). The average lifetime is $t_{\rm recycled,ave}=4.7$ 
Gyr, while median is $t_{\rm recycled,med}=4.6$ Gyr.}    
\label{fig.recyc}
\end{figure}

We follow the same process as above to generate a distribution of the radio lifetimes for 
the recycled pulsars found in NS-NS systems. However, these recycled pulsars in NS-NS 
systems have different spin period and magnetic field distributions compared to normal 
radio pulsars. Consequently, based on the observed pulsars in NS-NS systems, we adopt a 
uniform spin period distribution in the range $20~{\rm ms} < P_{\rm s} < 30~{\rm ms}$ and 
a log-normal magnetic field distribution with mean 
$\left< {\rm log}_{10}B {\rm (G)} \right> = 9.5$ and standard deviation 
$\sigma_{{\rm log}_{10}B} = 1$~\citep{Lorimer2008}. As opposed to normal radio 
pulsars, the Parkes Multibeam survey has only detected four recycled pulsars in NS-NS systems 
\citep{Manchester2005}. Thus, to account for this small number of detections, we repeat the 
process described above multiple times to ensure that there is no significant variation 
in the final lifetime distribution. The representative lifetime distribution for one 
these simulations is shown in Fig.~\ref{fig.recyc}.

\bibliographystyle{aa}

\end{document}